\documentclass[letterpaper,12pt]{article}
\usepackage{listings}
\usepackage{anysize}
\usepackage{bm}
\usepackage{graphicx}

\usepackage{amssymb}
\usepackage{xcolor}
\usepackage{threeparttable}
\usepackage{subfig}
\usepackage{url}
\usepackage{amsthm}
\usepackage{amsmath}
\usepackage{booktabs}
\usepackage{multirow}
\usepackage{times}
\usepackage[authoryear]{natbib}
\bibpunct{(}{)}{,}{a}{}{,}

\usepackage{setspace}

\newtheorem{theorem}{Theorem}

\usepackage[margin=1in]{geometry}

\newtheorem{proposition}[theorem]{Proposition}

\newtheorem{corollary}[theorem]{Corollary}

\newcommand{\be}{\begin{equation}}
	\newcommand{\en}{\end{equation}}
\newcommand{\ben}{\begin{equation*}}
	\newcommand{\enn}{\end{equation*}}
\newcommand{\bea}{\begin{eqnarray}}
	\newcommand{\ena}{\end{eqnarray}}

\title{Asset management with an ESG mandate}
\author{Michele Azzone\thanks{Department of Mathematics, Politecnico di Milano, Michele.azzone@polimi.it},
	\ Emilio Barucci\thanks{Department of Mathematics, Politecnico di Milano, Emilio.barucci@polimi.it},
	\ Davide Stocco\thanks{CREST, ENSAE, Institut Polytechnique de Paris, davide.stocco@ensae.fr}
}

\begin{document}
	\maketitle

	\date{\vspace{-3ex}}

	\vspace*{0.11truein}
	\begin{abstract}
		\noindent
We investigate the portfolio frontier and risk premia in equilibrium when institutional investors aim to minimize the tracking error variance under an ESG score mandate. If a negative ESG premium is priced in the market, this mandate can reduce portfolio inefficiency when the return over-performance target is limited.  In equilibrium, with asset managers endowed with an ESG mandate and mean-variance investors, a negative ESG premium arises. A result that is supported by empirical data. The negative ESG premium is due to the ESG constraint imposed on institutional investors and is not associated with a risk factor.
	\end{abstract}
	
	\vspace*{0.11truein}
	
	\vspace*{0.11truein}
	
	\vspace*{1cm}

	\newpage


	{
	\section{Introduction}

	Asset managers are facing the challenge of introducing ESG goals in their investment process.
	A request that comes from investors motivated by ESG themes and by regulation that has introduced the ESG classification of financial products
	and disclosure requirements to  enhance investors' awareness, e.g. see \cite{SEC,EC,FCA}.
	
	There are several approaches  to incorporate ESG themes in the asset management practice:
	positive/best-in-class screening, negative/exclusionary screening, ESG integration, impact investing, corporate engagement, sustainable themed investing.
	The ESG integration seems to be a promising approach, actually a large fraction of ESG funds follows this route, see \cite{alliance2018global}.
	ESG integration is not based on a 0-1 decision on the inclusion of an asset in the investment universe, it incorporates ESG themes into the portfolio construction.
	An easy way to implement it is to rely on ESG scores introducing a constraint such
	that the portfolio is characterized by a certain ESG score. We refer to this type of constraint as {\em ESG mandate},
	a feature of the fund that is often disclosed to investors.
	
	In this paper we address the consequences of introducing an ESG
	mandate for asset managers.
	Our analysis deals with mutual funds at benchmark rather than absolute return funds (e.g., hedge funds).
	Asset managers are worried by an ESG mandate as it reduces
	the investment universe, leading to under-diversification, and may conflict with a performance target in case of a
	trade-off between ESG score and return, see the debate on carbon or green premium \citep{BOLT,cao2023esg,CORN,duan2021carbon,liang2022responsible,Azzon}.
	The main result of our analysis is that the ESG mandate may not undermine
	the fiduciary duties of the asset manager provided that a negative ESG premium is priced; a relation established  in equilibrium
	and supported by the empirical evidence.
	
	The ESG mandate is introduced assuming that the asset manager minimizes the
	Tracking Error Variance (TEV, hereafter) with respect to a market benchmark, as in \cite{ROLL}, with the additional constraint of a  portfolio's weighted ESG score  greater than  the benchmark's one.
	Minimization of TEV comes from the remuneration of asset managers that is related to the relative (to the benchmark) performance.
	The framework builds upon the
	models proposed in \cite{PASTOR,PEDER}: the ESG score is provided by an ESG rating agency and the asset manager takes it as a datum,
its quality is not under scrutiny. This is the key difference with respect to \cite{AVRAM}, where ESG scores are treated as random variables. The ESG score is a feature of the asset and is not properly a risk factor.
	Differently from \cite{PASTOR,PEDER}, we do not introduce a preference for ESG in the investor's utility function but we include
	ESG considerations through a constraint on the weighted ESG score of the portfolio. We remark that the introduction of preference for ESG {\em à la}
	\cite{FAMA} is a viable approach to deal with ESG themes in asset management but has the limit that an estimation of the preference for ESG is needed. Instead, our approach closely mimics the practice in the asset management industry of including
	the ESG score as an additional  constraint on portfolio optimization without quantifying any ESG preference.
	
	The papers closest to ours are \cite{ANDER,BLITZ,BOLT2,SCH,DIZI,LING,RONC}.
	\cite{RONC} consider the construction of the portfolio frontier (minimizing the variance and not the TEV) adding an equality constraint on the ESG score or the carbon emission of the portfolio. \cite{BOLT2,SCH} consider the TEV or the variance minimization problem adding a constraint on the ESG score or on the GHG intensity, they show that the constraint does not necessarily lead to higher risk and lower return. \cite{ANDER,BOLT2} propose an asset management model based on the minimization of the TEV excluding assets with the worst carbon emission performance, they show that the additional constraint induces a limited risk increase, see also \cite{BLITZ,DIZI}.
	\cite{LING} introduce three types of investors: ESG-unaware, ESG-aware, ESG-motivate. The latter type maximizes a mean-TEV utility
	including a function increasing in the ESG score's excess with respect to the benchmark as in \cite{PEDER}.
		
  	We explore the connection among frontiers derived by minimizing portfolio variance, minimizing the TEV, and minimizing the
	TEV with the inclusion of an ESG mandate.
	We show that the integration of ESG themes into the
	asset management optimization problem may contribute to mitigate the inefficiency of portfolios constructed minimizing the TEV.
	Under some conditions, for a limited over-performance target relative to the benchmark, the ESG mandate renders
	a smaller variance for the portfolio frontier. Instead, when the performance target is high enough,
	the ESG mandate renders a higher variance.
	These results are obtained if the ESG mandate constraint is binding for a positive over-performance target
	relative to the benchmark. This occurs when a negative relationship between
	expected returns and ESG scores holds true and the
    benchmark is not too challenging (high return and high ESG score).
    	An ESG mandate plays a role similar to the one of a constraint on $\alpha$
	or  $\beta$ with respect to the benchmark, see \cite{BAPT,ROLL},
	or on total variance, see \cite{JOR}. The main result is that an ESG mandate does not undermine fiduciary duties of the
    asset manager yielding a mean-variance improvement. The root of the result is that a binding ESG mandate induces the asset manager
    targeting the TEV of the portfolio to over-invest in assets with a high Mean-Standard Deviation (M-SD) ratio, and therefore in assets with a lower variance.
    Thanks to the change of metrics (from TEV to variance), this feature renders a mean-variance improvement of the portfolio.

    The result is confirmed allowing the asset manager also to target an ESG over-performance with respect to the benchmark.
    Curiously enough, a more socially responsible investor is able to reach a significant mean-variance improvement for a limited return over-performance target. Instead, a less socially responsible investor reaches a small mean-variance improvement for a high return over-performance target. 
    	
	To test empirically the presence of a negative ESG premium, we expand the framework provided in \cite{BREN}, see
	\cite{BREN-LI, GOM} for empirical insights. We consider mean-variance retail investors and asset managers maximizing the mean-TEV utility with an ESG mandate. A
	negative ESG premium (lower returns associated to higher ESG scores) arises in
	equilibrium if the ESG constraint for the asset managers is binding for a positive over-performance target. 
	At the root of the result there is a rebalancing/demand channel rather than an ESG risk factor: the ESG constraint inflates the demand by institutional investors for ESG virtuous stocks.
Empirically, analyzing the US stock market, we provide positive evidence on this insight. This result adds to the literature on the presence of an ESG premium in financial markets, see e.g. \cite{ALESSI, AVRAM, BOLT, CORN,GORG,huynh2021climate, LIOUI,PEDER,PO,ZERBIB}. We are also able to verify that an ESG mandate leads to a mean-variance gain for a positive over-performance target, actually a mean-variance improvement is observed if the return over-performance target
	is between $0$ and $4.56\%$.
		
	
	The mean-variance improvement of portfolios obtained minimizing the TEV with an ESG mandate complements the analysis in \cite{LIND} where it is shown that ESG investing does not entail performance costs. The main insight of our analysis is that an ESG mandate does not  compromise the fiduciary duties of asset managers. A point made empirically in other papers, see \cite{ANDER,BLITZ,BOLT2,SCH}, the main novelty of ours is that we prove it analytically.
	
	The rest of the paper is organized as follows. In Section \ref{FRO}, we construct the TEV portfolio frontier with an ESG mandate. In Section \ref{PortAnalis}, we study the portfolio composition of the frontier.
	In Section \ref{MARK}, we address the market equilibrium analysis.
	In Section \ref{EMP}, we provide the empirical analysis.
	In Section \ref{EXT}, we extend the model also considering an ESG over-performance target.
	Section \ref{CONC} concludes. All proofs are in Appendix \ref{APP}.
	
	\section{Portfolio frontier with an ESG mandate}
	\label{FRO}
	
	Consider an economy with $N$ risky assets. Following best practices in the asset management industry and abstracting from agency problems between investors and managers,
	we assume that the asset manager minimizes the TEV conditional on a given level of expected return, while achieving an ESG score greater than  the  benchmark's one.

	We consider a benchmark $\mathbf{x_0} \in \mathbb{R}^N$ and
	a portfolio of the assets $\mathbf{x} \in \mathbb{R}^N$. We denote by $\boldsymbol{\mu} \in \mathbb{R}^N$ the vector of expected returns,
	$\boldsymbol{\xi} \in \mathbb{R}^N$  the vector of ESG scores,
	$\Omega \in \mathbb{R}^{N \times N}$ the returns variance-covariance matrix, and ${\bf 1} {\in \mathbb{R}^N}$  the unitary vector.
	$G$ is the
	expected return target of the asset manager relative to the benchmark: $G>0 \ (<0)$ means that the
	manager aims to over-perform (under-perform) the benchmark.
	
	The ESG mandate is introduced imposing the ESG score of the portfolio to be greater or equal than that of the benchmark. Therefore, the TEV frontier with an ESG mandate (hereinafter TEV ESG frontier) is obtained solving the following problem:
	\begin{align}
		\min_{\mathbf{x}}&\,(\mathbf{x}-\mathbf{x_0})^{\top}\Omega\, (\mathbf{x}-\mathbf{x_0})\\
		&\text{subject to} \nonumber\\
		&\;\mathbf{x}^{\top}{\bf 1}=1\\
		&(\mathbf{x}-\mathbf{x_0})^{\top}\boldsymbol{\xi} \ge 0 \label{CONST2}
		\\
		&(\mathbf{x}-\mathbf{x_0})^{\top}\boldsymbol{\mu} =G. \label{CONST3}
	\end{align}
	
	In what follows, we refer to the constraint (\ref{CONST2}) as binding if it holds as an equality.

	The portfolios of the frontier are identified in Proposition \ref{prop:frontier} in the Appendix.
		The vector $\mathbf{x}^*$ of portfolio weights of the TEV ESG frontier is
		\begin{equation}
			\mathbf{x}^*= \mathbf{x_0}-\frac{1}{2} \Omega^{-1}(\lambda_1 {\bf 1}+\lambda_2 \boldsymbol{\xi}+\lambda_3 \boldsymbol{\mu})
		\end{equation}
		where
		\begin{equation}
			\label{LAGR}
			\lambda_1=   \frac{ 2 (E \, A_E - A \, B_E)G }{ D_E} ,    \
			\lambda_2=  \frac{ 2 (A \, A_E - E \, C)G }{ D_E} \le 0,    \
			\lambda_3 = \frac{ 2 (  B_E \, C-A_E^2)G }{ D_E}     \
		\end{equation}
		if the constraint (\ref{CONST2}) is binding.

		If the constraint (\ref{CONST2}) is not binding, then the vector of portfolio weights is
		
		\begin{equation}
			\mathbf{x}^*= \mathbf{x_0}-\frac{1}{2} \Omega^{-1}(\hat{\lambda}_1{\bf 1}+\hat{\lambda}_3  \boldsymbol{\mu}) 	\label{COND}
		\end{equation}
		
		\begin{equation}
			\hat{\lambda}_1=   \frac{ 2 A \, G }{ D},     \
			\hat{\lambda}_3 = -\frac{ 2 C \, G }{ D},     \
			\label{eq:lagr_mult_roll}
		\end{equation}
		where,
		$$
		D_E = -2A \, E  \, A_E + A_E^2 \, B + A^2\,  B_E + E^2 \, C - B \,B_E  \, C, \ D = B\, C - A^2
		$$ and
		$$
		A={\bf 1}^{\top}\Omega^{-1}\boldsymbol{\mu}, \  \  B=\boldsymbol{\mu}^{\top}\Omega^{-1}\boldsymbol{\mu}, \  \   C={\bf 1}^{\top}\Omega^{-1}{\bf 1},
		$$
		$$
		A_E={\bf 1}^{\top}\Omega^{-1}\boldsymbol{\xi}, \  \  B_E=\boldsymbol{\xi}^{\top}\Omega^{-1}\boldsymbol{\xi}, \  \   E=\boldsymbol{\xi}^{\top}\Omega^{-1}\boldsymbol{\mu} \;\;.
		$$

	Notice that in case of a non binding ESG constraint, the portfolio coincides with the one in \cite{ROLL} and $\mathbf{x}^*$ is the rescaled difference between two portfolios:  $\frac{{\Omega}^{-1}{\bf 1}}{C}$ (Minimum Variance Portfolio, MVP) and $\frac{\Omega^{-1}\boldsymbol{\mu}}{A}$.
	
	As proved in Corollary \ref{cor:DE},
		for $G=0$, the ESG constraint (\ref{CONST2}) is always binding. For $G\neq 0$, it is binding in case
		\begin{equation}
			\label{CONSTR}
			\begin{cases}
				E-\frac{A}{C}A_E<0,\quad &G>0\\
				E-\frac{A}{C}A_E>0, \quad &G<0\;\;.
			\end{cases}
		\end{equation}	
	
Notice that the ESG constraint is always binding either for $G>0$ or $G<0$, depending on the data of the assets. If the constraint is binding, then $ D_E<0$.

	In what follows, we study the efficiency of the TEV ESG frontier. We compare  three frontiers in the mean-variance plane: the standard mean-variance portfolio frontier of \cite{MARCO},  the TEV portfolio frontier of \cite{ROLL} and the TEV ESG portfolio frontier.
	
	The variance of a portfolio belonging to the standard frontier is
	$$		
	Var_{Front}(G)=\frac{C}{D}\bigg(G+\mathbf{x_0}^{\top}\boldsymbol{\mu}-\frac{A}{C}\bigg)^2+\frac{1}{C}\;\;,
	$$
	where $G+\mathbf{x_0}^{\top}\boldsymbol{\mu}$ is the portfolio's expected return.
	
	The variance of the frontier obtained minimizing the TEV is
	\begin{equation}
	\label{VAR5}
		Var_{TEV}(G)= \mathbf{x_0}^{\top}\Omega \mathbf{x_0}-   ({\mathbf{x_0}}^{\top} ( \hat{\lambda}_1 {\bf 1} +  \hat{\lambda}_3  \boldsymbol{\bf{\mu}}))+\frac{CG^2}{D}\;\;.
	\end{equation}
	This frontier is obtained considering the case of a non binding ESG constraint for every $G$, see the portfolio in (\ref{COND}).
	
	The variance of the TEV ESG frontier with a binding ESG constraint for $G\ge 0$ is
	\begin{equation}
\label{VAR6}
		Var_{TEV_{ESG}} (G) = \begin{cases}
			\mathbf{x_0}^{\top}\Omega \mathbf{x_0}-   (\mathbf{x_0}^{\top} (\lambda_1 {\bf 1} + \lambda_2 \boldsymbol{\xi}   + \lambda_3  \boldsymbol{\mu})) + \frac{(A_E^2 - B_E C) G^2}  {D_E} \quad \quad &G\geq 0\\
			\mathbf{x_0}^{\top}\Omega \mathbf{x_{0}}-   ({\mathbf{x_{0}}}^{\top}( \hat{\lambda}_1 {\bf 1} +  \hat{\lambda}_3  \boldsymbol{\mu}))+\frac{CG^2}{D} &G<0
		\end{cases}\;\;.
	\end{equation}
	A specular expression for the TEV ESG frontier is obtained when the constraint is binding for $G \le 0$.
	
	By construction, the two frontiers $Var_{TEV}(G)$ and $Var_{TEV_{ESG}}(G)$ are on the right hand side
	of $Var_{Front}(G)$ in the mean-variance plane. While $Var_{TEV}$ dominates $Var_{TEV_{ESG}}$ in the mean-TEV plane,
	it is more complex to establish  their relationship in the mean-variance plane.
	
We observe that for a $G$ such that the ESG constraint (\ref{CONST2}) is not binding,
	the two  frontiers $Var_{TEV}(G)$ and $Var_{TEV_{ESG}}(G)$ coincide.
	For a $G$ such that the ESG constraint is binding, the two frontiers coincide if
	$$
	Var_{TEV}(G)-Var_{TEV_{ESG}}(G)=0,
	$$
	yielding the condition
	$$
	-   ({\bf x_{0}}^{\top} ( \hat{\lambda}_1 {\bf 1} +  \hat{\lambda}_3  {\bf \mu}))+\frac{CG^2}{D}+(x_0^{\top} (\lambda_1 {\bf 1} + \lambda_2 {\bf \xi}
	+ \lambda_3  {\bf \mu})) - \frac{(A_E^2 - B_E C) G^2}  {D_E}=0,
	$$
	which delivers up to two intersections $0$ and $G^*$, where
	\begin{align}G^*=\frac{2\mathbf{x_{0}}^{\top}\left((AD_E-D(EA_E-AB_E)){\bf 1}-D (AA_E-EC) \boldsymbol{\xi}   - (D_EC+D(B_EC-A_E^2))\boldsymbol{\mu} \right)}{D_E{C}-{D(A_E^2 - B_E C) } }.
		\label{G_star}
	\end{align}
	The second intersection is obtained if and only if the ESG constraint is binding for $G=G^*$. If this is not the case, then
	the two frontiers coincide in the non binding region while
	$Var_{TEV_{ESG}} (G)$ is dominated by $Var_{TEV}(G)$ in the binding one.
	In Figure \ref{figure:FrontierH}, we plot the three frontiers when the ESG constraint is binding for $G>0$ and either $G^*>0$ (on the left hand side) or $G^*<0$ (on the right hand side).
	In both cases, the two frontiers coincide for $G<0$. In the first case, there are some values of $G>0$ for which $Var_{TEV_{ESG}}(G)$ dominates $Var_{TEV}(G)$, in the latter, $Var_{TEV}(G)$  always dominates  $Var_{TEV_{ESG}}(G)$ for $G>0$.
	
	\begin{figure}[!htb]
		\centering
		\begin{minipage}{.5\textwidth}
			\centering
			\includegraphics[width=1\linewidth, height=0.25\textheight]{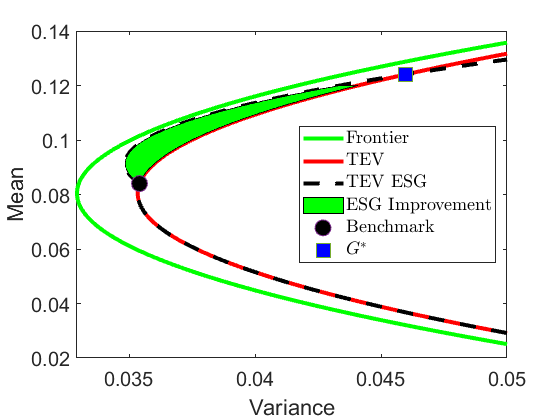}
		\end{minipage}%
		\begin{minipage}{0.5\textwidth}
			\centering
			\includegraphics[width=1\linewidth, height=0.25\textheight]{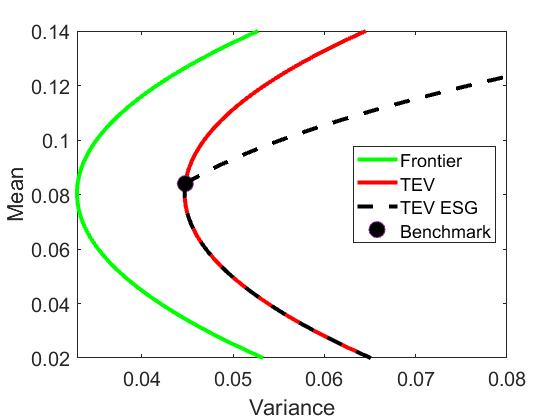}
		\end{minipage}
		\caption{The three frontiers when the constraint (\protect\ref{CONST2}) is  binding for $G>0$,
			$G^*>0$ on the left hand side and $G^*<0$ on the right hand side. The two figures are based on the assets considered in Section \ref{PortAnalis} and  two different benchmarks ($\frac{A}{C}=8\%$).}\label{figure:FrontierH}
	\end{figure}
	
	If the constraint (\ref{CONST2}) is binding for $G>0$ and $G^*>0$, then, for sufficiently large $G$,
	$Var_{TEV_{ESG}} (G)$ is dominated by $Var_{TEV}(G)$ as shown in Figure \ref{figure:FrontierH}. This can be proved analytically observing that
	the dominant term in the difference between the two variances, (\ref{VAR6}) and (\ref{VAR5}), for a large $G$ is the quadratic term:
	\[
	G^2\left(\frac{(A_E^2 - B_E C)}  {D_E}-\frac{C}{D}\right)=G^2\frac{(AA_E-EC)^2}{-DD_E}>0,
	\]
	which is positive because $D_E$ is negative and $D$ is positive.
	
	 We remark that, if the benchmark belongs to the mean-variance frontier, then the TEV ESG frontier intersects the mean-variance one only at the benchmark and is always dominated by the TEV frontier.
	
	In Figure \ref{figure:FrontierCloseFar}, we plot the three frontiers
	when the constraint is binding for $G>0$ and $G^*>0$.  We consider a benchmark with an expected return below $\frac{A}{C}$ (left hand side)
	and one above (right hand side).
	As shown analytically above, in both cases
	the $Var_{TEV_{ESG}} (G)$ frontier dominates the $Var_{TEV}(G)$ frontier for positive and low values of $G$ and the opposite holds true for a sufficiently large  $G$.
	In these cases, adding an ESG constraint to the minimization of TEV leads to a mean-variance improvement for a low positive value $G$.
	
	\begin{figure}[!htb]
		\centering
		\begin{minipage}{.5\textwidth}
			\centering
			\includegraphics[width=1\linewidth, height=0.25\textheight]{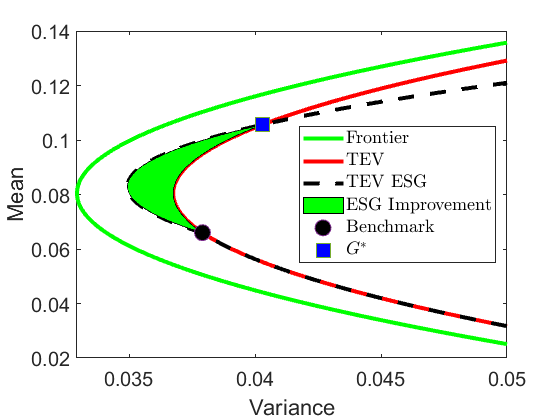}
		\end{minipage}%
		\begin{minipage}{0.5\textwidth}
			\centering
			\includegraphics[width=1\linewidth, height=0.25\textheight]{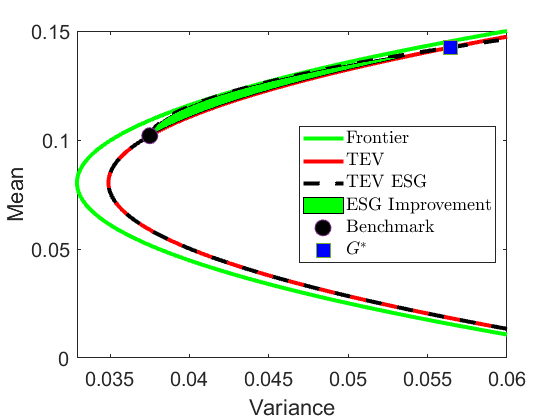}
		\end{minipage}
		\caption{The three frontiers when the constraint (\protect\ref{CONST2}) is  binding for $G>0$ and $G^*>0$.
			The two figures are based on the assets considered in Section \ref{PortAnalis} and  two different benchmarks ($\frac{A}{C}=8\%$). \protect\label{figure:FrontierCloseFar}}
	\end{figure}
	
	When the ESG constraint (\ref{CONST2}) is binding for negative $G$ the results are mirrored on the inefficient part of the frontier. The two frontiers coincide for $G\geq0$ and there is
	an intersection in $G=G^*$  if $G^*<0$.
	In Figure \ref{figure:FrontierIN}, we present an example in which the ESG constraint is binding for $G<0$ and $G^*<0$.
	The TEV ESG frontier intersects the TEV frontier in the non efficient region for $G=0$ and $G^*<0$.
	In this case, adding an ESG constraint to the minimization of TEV does not yield a mean-variance improvement.
	
	In \cite{PEDER}, the authors  conduct a similar analysis comparing the standard portfolio
	frontier, i.e., $Var_{Front}(G)$, with the one obtained adding an ESG mandate.
	They show that the two frontiers share a (tangent) portfolio.
	Our setting is different because the ESG constraint (\ref{CONST2}) concerns the ESG improvement
	with respect to the benchmark, and not the portfolio itself, moreover we minimize the TEV instead of the variance.
	
	\begin{center}
		\begin{figure}
			\centering
			\includegraphics[width=0.7\textwidth]{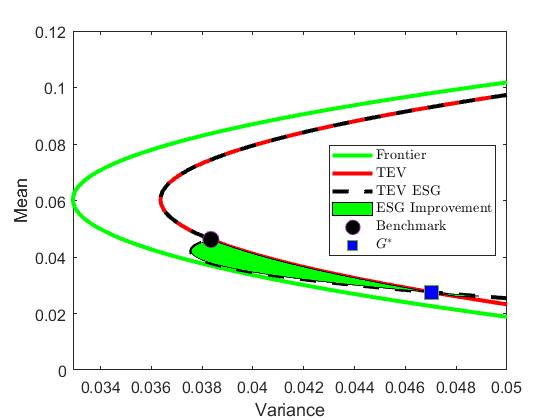}
			\caption{\small
				The three frontiers when the ESG constraint is binding for $G<0$ and $G^*<0$ ($\frac{A}{C}=6\%$).}		\label{figure:FrontierIN}
		\end{figure}
	\end{center}

	It is not straightforward to interpret
	condition (\ref{CONSTR}) that discriminates the values of $G$ for which the ESG constraint (\ref{CONST2}) is binding.
	In what follows, we discuss two particular cases focusing our attention on an ESG constraint binding for $G>0$.
	
	First, let us consider a diagonal $\Omega$ matrix, i.e., asset returns are not correlated. We denote the diagonal elements of the inverse matrix as $v_n=v_{n,n}, \ n=1, \dots, N$. Condition (\ref{CONSTR}) for a binding constraint for a positive $G$ reduces to
	\[
	{\sum_{i=1}^N  {\xi}_i v_i \left(\mu_i -\frac{A}{C}\right)  }<0\;\;.
	\]
	Hence, if the average of the expected returns of assets in excess of the expected return of the MVP
	weighted by the ESG scores divided by the return variances
	is negative, then the ESG constraint is binding for a positive $G$.
	Therefore, the ESG constraint is binding for a positive $G$ when the assets with high expected returns are characterized by low ESG scores and high variances.
	
	Second, let us consider a linear relation between $\xi$ and $\mu$, i.e., $\xi =\gamma \mu$. We can show that
	$$
	E-\frac{A}{C}A_E<0 \iff \gamma<0.
	$$	
	The ESG constraint (\ref{CONST2}) is binding for $G>0$ if and only if $\gamma <0$.
	The interpretation of this result is straightforward. If the asset manager targets an extra return with respect to the
	benchmark and the ESG score goes up with the expected return ($\gamma >0$) then the ESG mandate constraint (\ref{CONST2}) is not binding. A binding constraint is obtained when an inverse relation holds true ($\gamma <0$).
	
	Both these examples are aligned with the interpretation that the ESG mandate is binding if and only if a negative relationship between the ESG score and the expected return holds true.
	If this is the case, then a return and an ESG over-performance target are in conflict with each other
	but we have been able to show that a mean-variance improvement can be obtained for a limited over-performance target
	and, therefore, the ESG mandate does not undermine the fiduciary duties of asset managers towards investors.
	
	\section{TEV ESG portfolios}
	\label{PortAnalis}
	
	In what follows, we investigate the portfolio frontiers considering an illustrative example of four assets. We utilize these assets in all the examples on the frontier analysis in the rest of the paper, they were also considered in the examples of Section \ref{FRO}.
	In Table \ref{table:asset_improving}, we report the mean return, the M-SD ratio  and the ESG score of the assets.
	We also include the same pieces of information for the MVP and the two different benchmarks that were considered in Figure \ref{figure:FrontierCloseFar}.
	The two benchmarks lead to a different effect on the TEV frontier, we refer to the first one (on the left hand side in Figure \ref{figure:FrontierCloseFar}) as a {\em Risk Reducer} benchmark (B-RR), as it allows to reach a lower global variance compared to the TEV frontier, and to the second one as a {\em Return Enhancer} (B-RE).
	
	\begin{table}[h]
		\centering
		\begin{tabular}{ccccc|ccc}
			\toprule
			&A&B&C&D& B-RR&B-RE&MVP\\
			\hline
			Mean& 0.15&0.10&0.05&0.02&0.07&0.10&0.08\\
			M-SD ratio&		0.61&		0.45&		0.18&		0.08&0.34&0.53&0.44\\
			ESG score&0.07   & 0.10   & 0.17   & 0.67&0.29&0.16&0.31\\
			\bottomrule
		\end{tabular}
		\\[10pt]
		\caption{Mean, M-SD ratio and ESG score values of the four assets (A, B, C and D),  of the B-RR and B-RE portfolios and of the MVP.}
		\label{table:asset_improving}
	\end{table}
	The returns of the four assets are characterized by the following variance-covariance matrix
	
	\begin{equation*}
		\Omega =
		\begin{bmatrix}
			0.06&0.04&0.02&0.01\\
			0.04&0.05&0.03&0.02\\
			0.02&0.03&0.08&0.03\\
			0.01&0.02&0.03&0.06
		\end{bmatrix}\;\;.
	\end{equation*}

	\begin{center}
		\begin{figure}[h]
			\centering
			\includegraphics[width=0.7\textwidth]{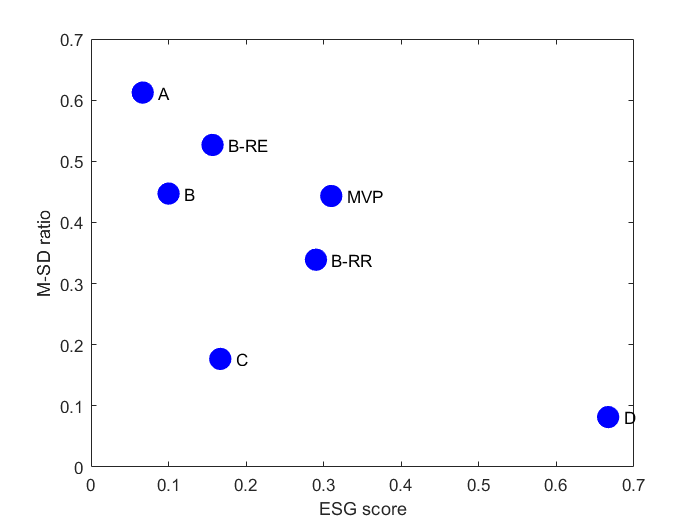}
			\caption{\small
				ESG score and M-SD ratio of the four assets (A, B, C, D), of the B-RE, B-RR portfolios and of the MVP.}		
			\label{figure:csi_sharpe_portfolio}
		\end{figure}
	\end{center}

In Figure \ref{figure:csi_sharpe_portfolio}, we plot the ESG score and the M-SD ratio of the four assets (A, B, C, D), of the B-RE, B-RR portfolios and of the MVP.
	Asset A is characterized by high M-SD ratio and low ESG score, while asset D shows low M-SD ratio and high ESG score. For these assets $E-\frac{A}{C}A_E=-1.92<0$, this entails that the ESG mandate is binding for a positive over-performance target. The four assets show a negative relationship between expected returns and ESG scores which is coherent with the discussions in Section \ref{FRO}. The B-RE benchmark is characterized by a higher M-SD ratio and a lower ESG score than the B-RR benchmark and the MVP.

In Figures \ref{figure:Portfolio_risk_red} and \ref{figure:Portfolio_RE}, on the left hand side, we plot the M-SD ratio-ESG score
combinations for portfolios belonging to the TEV frontier (red curve) and to the TEV ESG frontier (dashed black line), and, in the histograms on the right hand side, we show the
	portfolio weights of the MVP, TEV, TEV ESG portfolios for $G=1.5\%$, and B-RR, B-RE benchmark, respectively.
	In both cases, the over-performance target is smaller than $G^*$ and, therefore, the ESG constraint is binding and allows to reach a lower portfolio variance with respect to the minimization of TEV.
	For each asset, in bracket, we report the M-SD ratio difference with respect to the TEV portfolio frontier for the same ESG score.
	
	The region below the red curves in the pictures on the left hand side denotes the M-SD ratio-ESG score combinations that are dominated,
	in a M-SD ratio-ESG score sense, by the portfolios belonging to the TEV frontier while the region above the curves denotes combinations that dominate the TEV frontier.
	Above the red curve there are ESG score-M-SD ratio combinations that yield a ratio higher than that of the TEV portfolio frontier
	for a given ESG score. However, some of them are attainable through a feasible portfolio (they are on the right of mean-variance frontier) and when the ESG constraint is introduced, they can be achieved thanks to the fact that different metrics (TEV and variance) are considered, i.e., adding an additional constraint to a TEV minimization problem can lead to a better mean-variance outcome. This is shown by the dashed vertical line in correspondence of the two benchmarks, where the ESG constraint is binding,
	The highest point on the dashed line marks the maximum M-SD ratio attainable through a portfolio.
	
	Let us notice that assets A and D dominate the TEV frontier (positive numbers in brackets) while assets B and C are dominated (negative numbers in brackets). Comparing the weights of the assets (the first two bars for each asset), we observe  that a binding ESG mandate induces the institutional investor to over-invest in assets with higher M-SD ratio (lower variance and higher expected return) with respect to the simple TEV minimization: compared to the TEV frontier,
	TEV ESG portfolios are more concentrated on assets with a high M-SD ratio for a given ESG score, i.e., assets A and D.
	
Comparing the extension of the dashed black lines in the two figures, we observe that there is more space for a M-SD improvement in case of the
 B-RR benchmark than for the B-RE benchmark. This result confirms what is shown in Figure \ref{figure:FrontierCloseFar} and
 the  interpretation of this result is straightforward: if the benchmark is already characterized by a high M-SD ratio, then there is less space to  get a smaller variance thanks to the ESG mandate because portfolios are invested in assets with a high ratio.

\begin{figure}[!htb]
		\centering
		\begin{minipage}{.5\textwidth}
			\centering
			\includegraphics[width=1\linewidth, height=0.25\textheight]{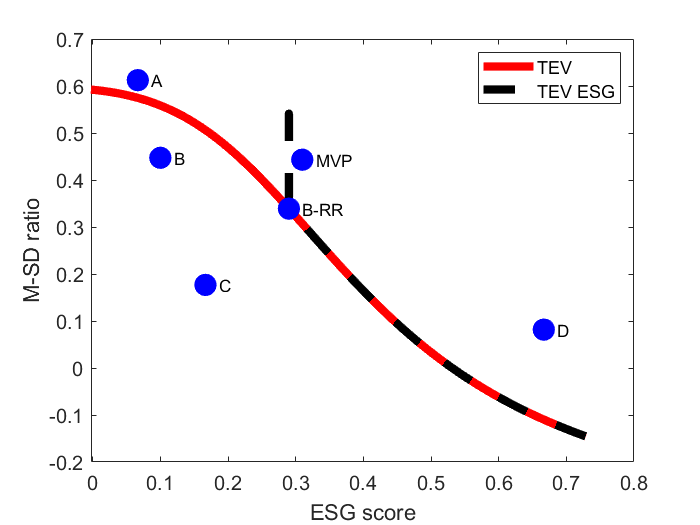}
		\end{minipage}%
		\begin{minipage}{0.5\textwidth}
			\centering
			\includegraphics[width=1\linewidth, height=0.25\textheight]{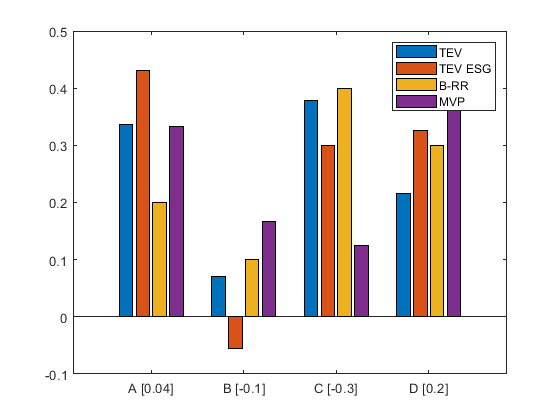}
		\end{minipage}
		\caption{\small
				M-SD ratio-ESG score combinations for portfolios belonging to the TEV frontier (red curve) and to the TEV ESG frontier (dashed black line) and portfolio weights of TEV, TEV ESG portfolios for $G=1.5\%$, MVP and the B-RR benchmark. For each asset, in bracket, we report the M-SD ratio difference with respect to the TEV portfolio with the same ESG score.}		
			\label{figure:Portfolio_risk_red}
				\end{figure}

\begin{figure}[!htb]
		\centering
		\begin{minipage}{.5\textwidth}
			\centering
			\includegraphics[width=1\linewidth, height=0.25\textheight]{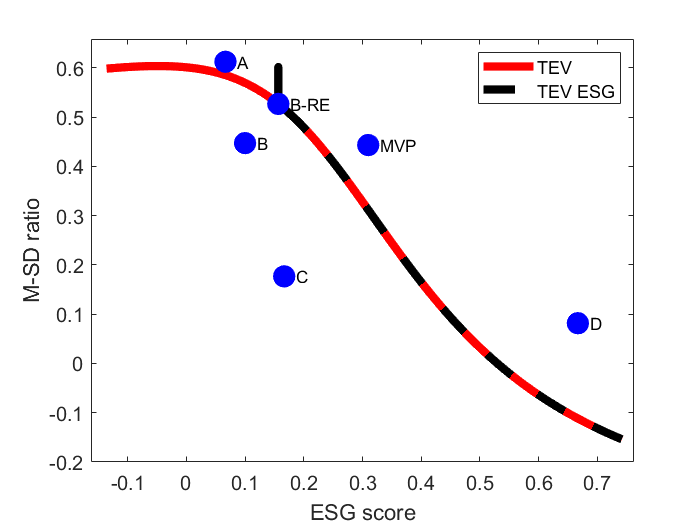}
		\end{minipage}%
		\begin{minipage}{0.5\textwidth}
			\centering
			\includegraphics[width=1\linewidth, height=0.25\textheight]{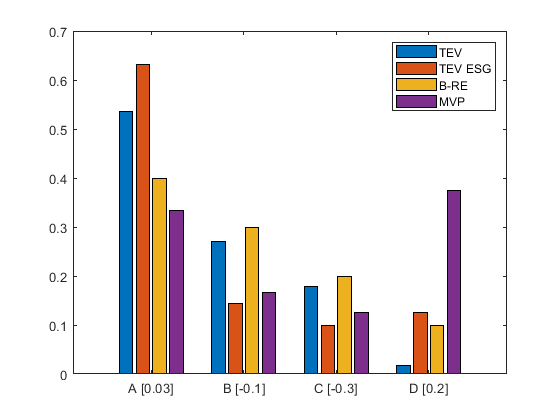}
		\end{minipage}
		\caption{\small
		M-SD ratio-ESG score combinations for portfolios belonging to the TEV frontier (red curve) and to the TEV ESG frontier (dashed black line) and the portfolio weights of TEV, TEV ESG portfolios for $G=1.5\%$, MVP and the B-RE benchmark. For each asset, in bracket, we report the M-SD ratio difference with respect to the TEV portfolio with the same ESG score.}		
			\label{figure:Portfolio_RE}
	\end{figure}

	In Section \ref{FRO}, we highlighted that for the same set of assets and different benchmarks, we can either observe
	a mean-variance improvement due to an ESG mandate or not, depending on $G^*$ that can be
	positive or negative for different benchmarks.
From equation \eqref{G_star},	we notice that
	\[
	G^* =  d_1 +d_2 \mathbf{x_{0}}^{\top} \boldsymbol{\mu} + d_3 \mathbf{x_{0}}^{\top} \boldsymbol{\xi}
	\]
	for some parameters $d_1$, $d_2$ and $d_3$, where $d_2, \ d_3 >0$.
	Consequently, the separation between positive and negative values of $G^*$ is defined by a straight line with a negative slope in the plane ESG score-expected return of the benchmark, see
	Figure \ref{figure:When_G_pos}, where the green area represents the combinations that yield a mean-variance improvement.
	The picture shows that a mean-variance improvement is attainable with an ESG mandate if the benchmark is not challenging both in terms of expected return and ESG score.

    These results go to the heart of the
	mean-variance improvement associated with an ESG mandate. A negative relation between ESG scores and expected returns
leads to a conflict between a return over-performance and an ESG target for the asset manager. However, this is not necessarily detrimental:
the ESG mandate leads to a mean-variance improvement if the expected return and the ESG score of the benchmark are not too high. If the benchmark's expected return is low enough (e.g. lower than that of the MVP), then the ESG mandate even leads to a reduction of the minimum variance attainable by minimizing the TEV, otherwise it contributes to reach higher returns. In both cases, the mean-variance improvement is due to the fact that a binding ESG constraint leads the asset manager to over-invest in assets with higher M-SD ratios yielding a mean-variance improvement.

	\begin{center}
		\begin{figure}[h]
			\centering
			\includegraphics[width=0.7\textwidth]{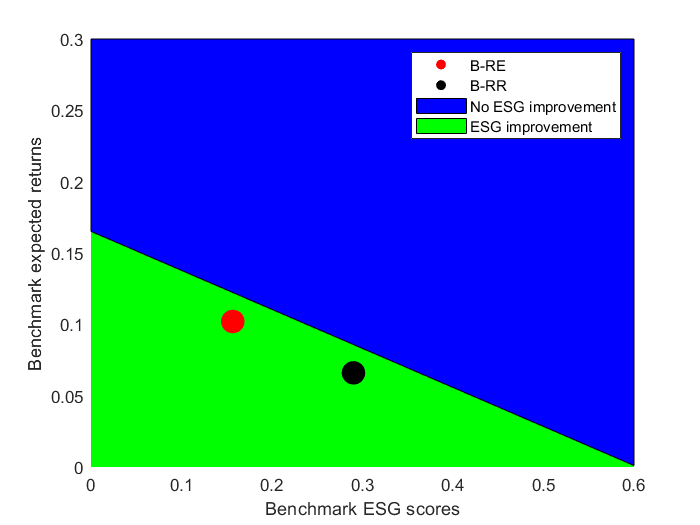}
			\caption{\small
				Region of mean-variance improvement of an ESG mandate (green area) in the plane ESG score-mean return of the benchmark.
				B-RE and B-RR portfolios are represented by the red and the black dot, respectively.}		
			\label{figure:When_G_pos}
		\end{figure}
	\end{center}

	\section{Market equilibrium}
	\label{MARK}
	A mean-variance improvement with an ESG mandate 
	is obtained if a negative premium is priced by the market. To test this hypothesis, we consider the market model in \cite{BREN,BREN-LI} introducing an ESG mandate for institutional investors.
	
	We consider an economy with $N$ risky assets and a risk-free asset with return $r_f$.  Retail investors maximize a mean-variance utility
	and can also invest in the risk-free asset. Institutional investors are remunerated according to their performance relative to the benchmark, and therefore they maximize a mean-variance utility defined with respect to the tracking error (mean-TEV utility) and can only invest in risky assets. We keep the number and the wealth of retail and institutional investors constant.
	
	The $i^{th}$ institutional investor ($i=1,\dots, I)$
	maximizes the mean-TEV utility defined with respect to the benchmark $\mathbf{x_{0i}}$ subject to the ESG mandate:
	\begin{align}
		\max_{\mathbf{x_i}}\,&(\mathbf{x_i}-\mathbf{x_{0i}})^{\top}\boldsymbol{\mu}-\frac{a_i}{2} (\mathbf{x_i}-\mathbf{x_{0i}})^{\top}\Omega\,(\mathbf{x_i}-\mathbf{x_{0i}}) \label{OPTIM_INST}\\
		&\text{subject to} \nonumber\\
		&\;\mathbf{x_i}^{\top}{\bf 1}=1 \label{CONSTR1}\\
		&(\mathbf{x_i}-\mathbf{x_{0i}})^{\top}\boldsymbol{\xi} \ge 0 \label{CONSTR2}\;\;,
	\end{align}
	where $a_i$ is the coefficient of absolute risk aversion.
	The constraint (\ref{CONSTR2}) is the key novelty with respect to the framework in \cite{BREN,BREN-LI}.
	The analysis can be easily extended to a setting with heterogeneous ESG over-performance targets.
As proved in Proposition \ref{prop:eq_inv}, the optimal portfolio for the institutional investor is
		\begin{equation}
			\mathbf{x_i}^*= \mathbf{x_{0i}}+\frac{1}{a_i} {\Omega}^{-1}( {\boldsymbol{\mu}}-\omega_1\mathbf{1}-\omega_2\boldsymbol{\xi}) \label{x_inst}\;\;,
		\end{equation}
		where \begin{equation*}
			\begin{cases}
				\omega_1=\frac{A}{C}-\omega_2 Z,\quad \omega_2=\frac{\left(\boldsymbol{\mu}-\frac{A}{C}\mathbf{1}\right)^{\top}\Omega^{-1}\boldsymbol{\xi}}{\left(\boldsymbol{\xi}-Z\mathbf{1}\right)^{\top}\Omega^{-1}\boldsymbol{\xi}}\quad \quad &\text{if (\ref{CONSTR2}) is binding}\\
				\omega_1=\frac{A}{C}, \quad \omega_2=0 \quad \quad &\text{otherwise}
			\end{cases}
		\end{equation*}
		and $Z = \frac{\boldsymbol{\xi}^{\top}{\Omega}^{-1}\mathbf{1}}{\mathbf{1}^{\top}{\Omega}^{-1}\mathbf{1}}$.
		
	Moreover,	the ESG constraint  (\ref{CONSTR2}) is binding if and only if
		\begin{equation}
			\label{COND4}
			E-\frac{A}{C}A_E<0\;\;.
		\end{equation}

	As expected, the ESG constraint is binding if and only condition \eqref{CONSTR} is verified for $G > 0$. As a matter of fact, the portfolio of the mean-TEV optimizer belongs to the efficient part
of the TEV ESG frontier which is different from the TEV frontier only if the ESG constraint is binding for $G>0$.
	
	The optimal portfolio of the institutional investor in (\ref{x_inst}) corresponds to
	the one in \cite{BREN-LI} and does not depend on the ESG score of the assets if the ESG constraint is not binding.
	Instead, if the ESG constraint is binding, then the optimal portfolio belongs to the TEV ESG frontier built
	in the previous section with $\omega_1$ and $\omega_2$ dependent on the ESG score.
	
	The
	$l^{th}$ retail investor ($l=1, \dots L$) does not take into account the ESG score and maximizes the standard mean-variance utility function:
	\begin{align}
		\max_{\boldsymbol{y_l}}&\quad \mathbf{y}_l^{\top}(\boldsymbol{\mu}-r_f\mathbf{1})-\frac{a_l}{2} \mathbf{y_l}^{\top}\Omega\mathbf{y}_l\;\;,
	\end{align}
	where $a_l$ is the coefficient of absolute risk aversion. The problem is equivalent to the one considered in \cite{BREN-LI} and
	the optimal portfolio is given by:
	\begin{equation}
		\mathbf{  y^*_l}=\frac{1}{a_l}\Omega^{-1}(\boldsymbol{\mu}-r_f\boldsymbol{1})\;\;\label{x_ind}.
	\end{equation}
	
	Let $W_i$ and $W_l$ the exogenous wealth of the $i^{th}$ institutional investor and of the $l^{th}$ individual investor, then
	in equilibrium we have:
	\begin{equation}
		\label{EQUI}
		W_m\mathbf{x_m}=\sum_{i=1}^I W_i \mathbf{x_i}+\sum_{l=1}^L W_l\mathbf{ y_l}\;\;,
	\end{equation}
	where $W_m$ is the market wealth ($W_m=\sum_{i=1}^I W_i+\sum_{l=1}^LW_l$) and  $\mathbf{x_m}$ is the market portfolio.

The following result can be obtained for the expected returns of the assets in equilibrium. If the ESG constraint is binding, then
		\begin{equation}
			\boldsymbol{\mu}={r_f^*}\mathbf{1}+\theta_1 \Omega \mathbf{x_m}-\theta_2\Omega \mathbf{x_0}-{\Gamma}\boldsymbol{\xi}\;\;, \label{eq:Eq}
		\end{equation}
		where $r_f^*$, $\theta_1$, $\theta_2$ and $\Gamma$ are constants defined as follows:
		
		\begin{itemize}
			\item[i.] $r_f^*= \frac{\left(\omega_1\sum_{i=1}^I \frac{W_i}{a_i}+\sum_{l=1}^L \frac{W_l}{a_l}r_f\right)}{\delta}$
			\item[ii.] $\theta_1 = \frac{W_m}{\delta}$
			\item[iii.] $\theta_2 = \frac{ \sum_{i=1}^I W_i}{\delta}$
			\item[iv.] $\Gamma = -\frac{\sum_{i=1}^I\frac{W_i}{a_i}\omega_2}{\delta}$
		\end{itemize}
		
		and $\delta = \sum_{i=1}^I \frac{W_i}{a_i}+ \sum_{l=1}^L \frac{W_l}{a_l}$.
For the proof see Proposition  \ref{prop:equilibrium_tot}.

	The beta of asset $j$ with respect to the market is
	$\beta_{mj} := \frac{ Cov(r_m, r_j)}{Var(r_m)} = \Omega_{j}\mathbf{x_m}$, where $r_m$ is the return of the market portfolio,
	$r_j$ is the return of asset $j$ and $\Omega_{j}$ is the $j^{th}$ row of matrix $\Omega$.
	The beta of asset $j$ with the benchmark is $\beta_{bj} := \frac{ Cov(r_{be}, r_j)}{Var(r_{be})} = \Omega_{j}\mathbf{x_{0}}$, where $r_{be}$ is the benchmark return.
	Hence, for  asset $j$ it holds
	\begin{equation}
		\label{EQT}
		\mu_j= r_f^*+\theta_1^*\beta_{mj}\mu_m-\theta_2^*\beta_{bj}\mu_{be}-\Gamma \xi_j, \;\; j=1, \dots, N,
	\end{equation}
	where $\theta_1^* = \theta_1 \sigma_m^2$, $\theta_2^* = \theta_2 \sigma_{be}^2$, $\sigma_m^2$ and $\sigma_{be}^2$ are the
	market and the benchmark variance, respectively.
	
	Notice that the ESG score of asset $j$ enters the expected return through the term $-\Gamma \xi_j$.
	$\Gamma$ is positive because $\omega_2$, the Lagrange multiplier associated to the binding ESG constraint, is negative.
	
		A negative ESG premium has been already established in the literature, see \cite{AVRAM,PASTOR}, the main novelty of our analysis concerns
		 the mechanism which is not due to a preference for ESG assets by investors but to the constraint for asset managers provided by the ESG mandate which leads to over-investment in virtuous stocks.
	The negative ESG premium arises in equilibrium when the ESG constraint is binding for institutional investors maximizing a mean-TEV utility and facing retail mean-variance investors in the market. Actually, in Section
\ref{FRO} we have shown that the constraint is binding if market data exhibit a negative relationship between ESG score and return.
Consequently, this equilibrium result and the condition for a binding constraint complement each other.

	\section{Empirical analysis}
	\label{EMP}
	
	In what follows, we provide an empirical analysis focusing on the US market.
	The data set consists of the stocks belonging to the Russell3000 index at least one year in the $2017-2022$ period.
	We consider monthly stock total returns and collect monthly ESG scores from Refinitiv. The Refinitiv ESG score is a widely used sustainability rating system, it covers most of the  Russell3000 index market capitalization starting from 2017, see \cite{Refinitiv}.
	In  Table \ref{table:Data left}, we provide the summary statistics of the stocks of the Russell3000 endowed with an ESG score and, therefore, included in the sample for each year (from January 2017 to December 2022).
	We remove from the analysis the stocks that have no ESG score in any year of the sample.
	Although the coverage of stocks with an ESG score is already 80\% of the market capitalization in 2017, the coverage 
	increases over time reaching more  than 70\% of Russell3000 stocks from 2019 onward.
	
	At each point in time,  we consider the monthly stock return and the latest available ESG score.
	ESG scores are published on an annual basis but on different dates. In our monthly time-series  we consider a
	constant ESG score from one release to the next one.
	If an ESG score is missing for a given month we replace it with the previous observation.
	Similarly to  \cite{DELV, SERA, LIO22},
	we normalize the ESG scores: for each month, we
	subtract the average ESG score of all the stocks
	from each stock's ESG score.

	\begin{table}[]
		\centering
		\begin{tabular}{|c|c|c|c|}
			\hline Year & \# stocks &     \% stocks &              \% market capitalization \\
			\hline \hline
			2017 &            954 &          32.23 \% &  84.96 \% \\  2018 &            1575 &             52.29 \% &    92.28 \% \\ 2019 &            2145 &             71.69 \% &   95.46 \% \\2020 &            2422 &         79.18 \% &  94.16 \% \\ 2021 &            2294 &        74.84 \% &  94.35 \% \\ 2022 &            2159 &           72.86 \% &  95.03 \% \\
			\hline
		\end{tabular}
		\\[10pt]
		\caption{Number of stocks with available Refinitiv  ESG score, fraction of the Russell3000
			stocks with available Refinitv ESG score, fraction of the market capitalization of the
			Russell3000 stocks covered by the data set.}
		\label{table:Data left}
	\end{table}

	We notice high volatility and outliers in February 2020, corresponding to the onset of the global COVID-19 pandemic, and in February 2022 in correspondence of the Russian invasion of Ukraine.
	To tackle this issue, we perform a winsorization on returns, which limits extreme values in the data set.
	We adopt a 95\% winsorization which sets all observations below the 2.5th percentile equal to the 2.5th percentile, and all observations
	above the 97.5th percentile equal to the 97.5th percentile.
	Following this procedure, we do not eliminate outliers but we limit their effect, for more details on the technique see \cite{Bali}.

	The first step in our empirical analysis concerns
	the equilibrium model introduced in Section \ref{MARK}.
	The main prediction of the model is a negative ESG risk premium in equilibrium if and only if there are asset
	managers with a positive over-performance target and the ESG constraint is binding for it.
	To test the equilibrium model in \eqref{eq:Eq}, we follow a procedure similar to the one in \cite{BREN-LI}.
	
	First, we estimate the residuals from the regression of the benchmark return on the market return:
	$$
	r_{be}(t)-r_f=\beta_{mb}(r_{m}(t)-r_f)+ e(t),
	$$
	where $r_{be}(t)$ and $r_m(t)$ are the return of the benchmark and of the market, respectively.
	Consistently with the analysis of the TEV ESG frontier conducted above, we proxy the market return with the weighted average of the returns of the stocks in our data set and
	the benchmark return with the weighted average of the returns of the 500 highest capitalized
	stocks.
	The estimated beta is $99\%$ which is consistent with the results obtained in \cite{BREN-LI}.

	As a 
second step, for each stock $j$ we compute the betas of the stock with the market return and the
	residuals from the above regression: $\beta_{mj}$ as defined in Section \ref{MARK} and $\beta_{ej}= \frac{Cov(e, r_j)}{Var(e)}$, respectively.
	As observed in \cite{BREN-LI}, we notice that
	\begin{align*}
		\beta_{bj}
		&=\frac{Cov(r_{be}(t),r_j(t))}{Var(r_{be}(t))}=\frac{\beta_{mb}Cov(r_{m}(t),r_j(t))+{Cov(e(t),r_j(t))}}{Var(r_{be}(t))}\\&=\beta_{mb}\beta_{mj}\frac{Var(r_m(t))}{Var(r_{be}(t))}+\beta_{ej}\frac{Var(e(t))}{Var(r_{be}(t))},
	\end{align*}
	then, we substitute this expression in the equilibrium model \eqref{EQT} obtaining
	\begin{equation}
		\label{EQW}
		{\mu}_j= a+b_1\beta_{mj}-b_2\beta_{ej}-\Gamma \xi_j, j=1, \dots, N,
	\end{equation}
	where $a$, $b_1$, $b_2$ and $\Gamma$ are positive constants.
	At this point, we estimate a cross-sectional regression on the average returns of the stocks in our data set to verify whether the equilibrium model, and in particular the negative ESG risk premium, is supported by market data.
	In Table \ref{table:Regressions}, we report the estimated parameters.
	We repeat the analysis including  75\%, 70\%, 60\%, 40\% and 20\% of stocks sorted by market capitalization from our data set.
	In all cases, the estimated $\hat{\Gamma}$ is positive, statistically significant at $1\%$, the $R^2$ is close to 10\% and goes up as we restrict our attention to well capitalized stocks.
	Notice that $\hat{b}_2$ is negative, in \cite{BREN-LI} a positive or negative coefficient is obtained depending on the period and on the size of companies.

	\begin{table}
		\centering
		\begin{threeparttable}
			\begin{tabular}{lccccc}
				\toprule
				& 20\% & 40\%& 60\%& 70\%& 75\%\\
				\midrule
				\multirow{2}{*}{$\hat{a}$}&0.4836$^{***}$ & 0.6199$^{***}$ & 0.5512$^{***}$ & 0.5398$^{***}$ & 0.5134$^{***}$ \\
				& (0.1253) & (0.0859) & (0.0729) & (0.0679) & (0.0656) \\
				\multirow{2}{*}{$\hat{b_1}$}&1.001$^{***}$ & 0.7837$^{***}$ & 0.7257$^{***}$ & 0.6751$^{***}$ & 0.6568$^{***}$ \\
				& (0.1186) & (0.0841) & (0.0713) & (0.0661) & (0.0639) \\
				\multirow{2}{*}{$\hat{b_2}$}&-0.0522$^{***}$ & -0.0502$^{***}$ & -0.0499$^{***}$ & -0.0486$^{***}$ & -0.0474$^{***}$ \\
				& (0.0082) & (0.0058) & (0.0048) & (0.0046) & (0.0044) \\
				\multirow{2}{*}{$\hat{\Gamma}$}&1.2034$^{***}$ & 1.2881$^{***}$ & 1.0842$^{***}$ & 0.9477$^{***}$ & 0.8385$^{***}$ \\
				& (0.2309) & (0.1585) & (0.1395) & (0.1321) & (0.1277) \\
				\midrule
				$R^2_{adj}$&16.4&12.8&9.77&8.32&7.71\\
				\bottomrule
			\end{tabular}
			\begin{tablenotes}
				\scriptsize
				\item \leavevmode\kern-\scriptspace\kern-\labelsep Note:$^{*}$p$<$0.10,$^{**}$p$<$0.05,$^{***}$p$<$0.01
			\end{tablenotes}
		\end{threeparttable}
		\caption{Coefficients of the cross-sectional regression (\ref{EQW})
			including the top  20\%, 40\%, 60\%, 70\% and 75\% stocks sorted by market capitalization in the data set.}
		\label{table:Regressions}
	\end{table}
	
	The equilibrium model with asset managers and an ESG mandate in Section \ref{MARK}
	provides a straightforward implication: a negative ESG risk premium is obtained if and only the ESG mandate is binding for a positive
return over-performance target. A condition that is ensured precisely by a negative relationship between expected returns and ESG scores.
	Therefore, the negative risk premium obtained in the cross-sectional regression provides positive evidence on the joint hypothesis of the
	presence of asset managers targeting a mean-TEV utility and of a binding ESG mandate with a positive over-performance target.
	
	To further evaluate the relevance of the ESG factor, we consider the CAPM cross-sectionally:	
	\begin{equation}
		\label{EQWCAPM}
		\mu_j= a+b_1\beta_{mj},\ \  j=1, \dots, N,
	\end{equation}
	and the five factors Fama and French model, see \cite{FAMA}, with and without the ESG component following the approach used to estimate the regression in
	(\ref{EQW}):
	\begin{equation}
		\label{EQWFFESGTEV}
		\mu_j= a+b_1\beta_{mj}+b_{SMB} SMB + b_{HML} HML + b_{RMW} RMW + b_{CMA} CMA-b_2\beta_{ej}-\Gamma \xi_j,
	\end{equation}
	where $SMB$ (Small Minus Big) is the average return on nine small-minus-big stock portfolios, $HML$ (High Minus Low) is the average return on two value-minus-growth portfolios,
	$RMW$ (Robust Minus Weak) is the average return on two robust-minus-weak operating profitability portfolios and
	$CMA$ (Conservative Minus Aggressive) is the average return on two conservative-minus-aggressive investment portfolios.
	
	Results of the regressions are reported in Table \ref{table:Regressions2} for top 75\% of stocks sorted by market capitalization in the data set.
	We report the results for the CAPM, Fama and French 3 factors model, i.e., only the first three factors (FF3), Fama and French 5 factors model (FF5),
	the model in \cite{BREN-LI} (TEV), a model only including ESG score (ESG), the CAPM including also the ESG score (CAPM + ESG),
the TEV model (TEV), the TEV model considering also the ESG score (TEV + ESG) and the Fama and French model with TEV investors and ESG score (FF5+TEV ESG).
TEV+ESG model corresponds to column 75\% in Table \ref{table:Regressions}.
	
	Comparing the TEV and the CAPM regressions, we observe that the model with institutional investors targeting a benchmark significantly improves the regression in terms of $R^2$. Adding an ESG factor
    (TEV+ESG), we observe a further improvement in the regression performance, the $R^2$ goes form $1.84$ to $5.78$ and then to $7.71$.
	This result highlights that the model proposed in our analysis is well grounded from an empirical point of view.
	
	Considering the Fama and French model we do confirm the positive sign for $\Gamma$, and therefore that a negative ESG premium is priced by the market, but the $R^2$ improvement is limited. Notice that we are not able to provide a theoretical model incorporating TEV investors and the ESG component in the multifactor model, and therefore
	we may interpret this evidence as a robustness check of the main result of the analysis centered on the CAPM.
	
	To provide an illustrative evidence of the presence of a negative ESG premium, in
	Figure \ref{figure:ESG_MU}, we represent the scatter plot of average ESG score and average monthly return over the full time-interval
	of the 20\% largest stocks by market capitalization and the regression line of the ESG model in Table \ref{table:Regressions2} (ESG column).
	The sample includes 529 stocks and is a good proxy of the S\&P 500 index.
	The estimated coefficient is negative consistently with a negative ESG risk premium.

	\begin{table}
		\centering
		\begin{threeparttable}
			\resizebox{\textwidth}{!}{\begin{tabular}{lcccccccc}
					\toprule
					& CAPM& FF3& FF5& TEV & ESG& CAPM+ESG&TEV + ESG&FF5+TEV ESG\\
					\midrule
					\multirow{2}{*}{$\hat{a}$}&0.5067$^{***}$ & 0.4941$^{***}$ & 0.5336$^{***}$ & 0.4918$^{***}$ & 0.9233$^{***}$ & 0.521$^{***}$ & 0.5134$^{***}$ & 0.5438$^{***}$ \\
					& (0.0675) & (0.0616) & (0.0611) & (0.0662) & (0.0231) & (0.0674) & (0.0656) & (0.0609) \\
					\multirow{2}{*}{$\hat{b}_1$}&0.3759$^{***}$ & 0.572$^{***}$ & 0.4349$^{***}$ & 0.6058$^{***}$ &  & 0.381$^{***}$ & 0.6568$^{***}$ & 0.643$^{***}$ \\
					& (0.0603) & (0.0681) & (0.0807) & (0.0641) &  & (0.0601) & (0.0639) & (0.0925) \\
					\multirow{2}{*}{$\hat{b}_{SMB}$}& &0.0765$^{*}$ & -0.0667 & &  &&  & -0.4761$^{***}$ \\
					&  &(0.0413) & (0.0599) &  &  & &  & (0.116) \\
					\multirow{2}{*}{$\hat{b}_{HML}$}& &-0.8556$^{***}$ & -0.2612$^{***}$ &  & & && -0.3572$^{***}$ \\
					&  &(0.0453) & (0.0989) &  & & & & (0.1035) \\
					\multirow{2}{*}{$\hat{b}_{RMW}$}& & &-0.0684 &  & & & &-0.0778$^{*}$ \\
					&  & &(0.0448) &  & & & &(0.0446) \\
					\multirow{2}{*}{$\hat{b}_{CMA}$}& & &-0.4268$^{***}$ &  & & & &-0.4152$^{***}$ \\
					&  & &(0.0592) &  & & & &(0.0591) \\
					\multirow{2}{*}{$\hat{b}_2$}& & & &-0.04$^{***}$  & & &-0.0474$^{***}$ &0.048$^{***}$ \\
					&  & & &(0.0043) & & &(0.0044) &(0.0131) \\
					\multirow{2}{*}{$\hat{\Gamma}$}& & & & &0.4749$^{***}$ &0.4924$^{***}$  &0.8385$^{***}$  &0.341$^{***}$ \\
					&  & & & & (0.1281)& (0.1269)& (0.1277)&(0.1229) \\
					\midrule
					$R^2_{Adj}$&1.84&18.5&20.5&5.78&0.63&2.52&7.71&21.2\\
					AIC&5760.9&5387.2&5339.7&5679.1&5785.8&5747.9&5638.3&5327.7\\
					\bottomrule
			\end{tabular} }
			\begin{tablenotes}
				\scriptsize
				\item \leavevmode\kern-\scriptspace\kern-\labelsep Note:$^{*}$p$<$0.10,$^{**}$p$<$0.05,$^{***}$p$<$0.01
			\end{tablenotes}
		\end{threeparttable}
		\caption{Coefficients of the cross-sectional regressions (\ref{EQWCAPM}) and (\ref{EQWFFESGTEV})
			including the top 75\% stocks sorted by market capitalization in the data set. We report the results for the CAPM,
        Fama and French 3 factors model (FF3),	Fama and French 5 factors model (FF5), the model with institutional investors (TEV), a model including only the ESG score (ESG), the CAPM with ESG score (CAPM + ESG), the TEV model with ESG mandate (TEV + ESG) and the
        model including the risk factors of Fama and French, TEV and ESG score (FF5+TEV ESG).}
		\label{table:Regressions2}
	\end{table}

	\begin{figure}
		\centering
		\includegraphics[width=0.6\textwidth]{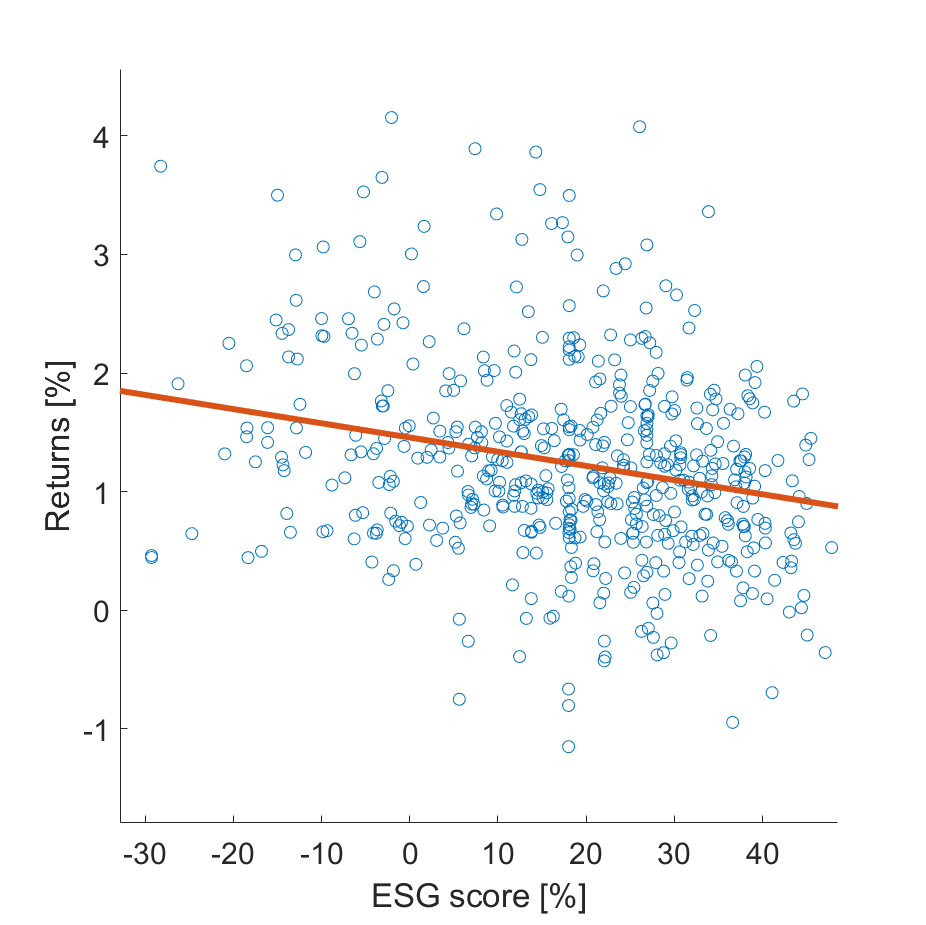}
		\caption{\small
			Average ESG score and average return over the time-interval of the 20\% top capitalized stocks (dots) and the estimated ESG regression in Table \ref{table:Regressions2} (continuous line).}		\label{figure:ESG_MU}
	\end{figure}

	The second step of our analysis consists in verifying whether the ESG constraint is binding for a positive or a negative $G$.
	We need to check wether condition \eqref{CONSTR} is verified for $G> 0$ or $G < 0$.
	Given the very large sample size of the data set, we cannot estimate
	the variance-covariance matrix for the full sample of stocks.
	Hence, we consider  eleven equally weighted portfolios based on the sectors  of
	the Refinitive Business Classification (TRBC sectors).
	We discard the stocks with low market capitalization that present also sparser ESG data, we only consider the 75\% highest capitalized stocks.
	As a benchmark, we consider the value weighted portfolio composed by the 500 highest capitalized
	stocks in December 2022.
	As far as the ESG score is concerned, we consider its average
	over the sample period.
	The computation renders
	$$
	E-\frac{A}{C} A_E=-0.27
	$$
	and, therefore, the constraint \eqref{CONSTR} is binding for $G> 0$.
	We also obtain $G^*=4.56\%>0$,
	and therefore, the $Var_{TEV_{ESG}}$ dominates the $Var_{TEV}$
	for $G \in (0, 4.56\%)$, while the opposite holds true for a $G > G^*$. 	
	In  Figure \ref{figure:VAR_IMPROV}, we report the  variance difference
	between $Var_{TEV_{ESG}}$ and $Var_{TEV}$ varying $G\ge 0$.
	The risk improvement associated with an ESG mandate amounts up to
	1\% of  the variance of the MVP.
	
	\begin{figure}
		\centering
		\includegraphics[width=0.6\textwidth]{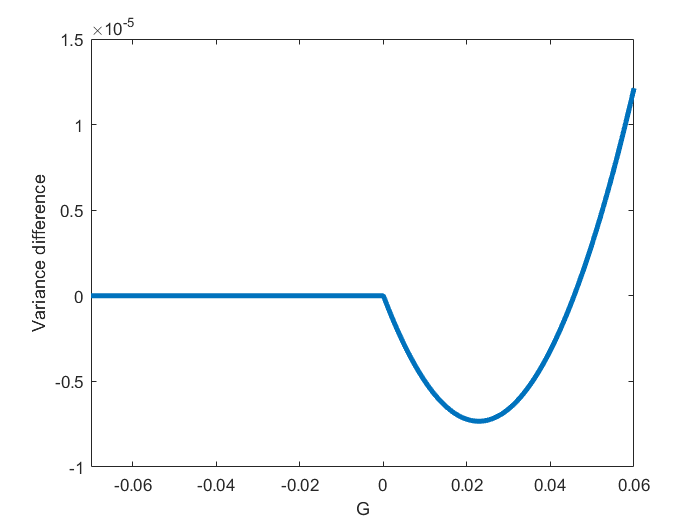}
		\caption{\small
			$Var_{TEV_{ESG}} - Var_{TEV}$ for the eleven equally weighted portfolios based on the sectors  of
			the Refinitive Business Classification (TRBC sectors).}		\label{figure:VAR_IMPROV}
	\end{figure}

	As a robustness check, we perform the analysis including only the top
	70\%, 60\%, 40\% and 20\% stocks of the data set, sorted according to the market capitalization.
	For the sake of consistency, when computing  portfolios based on the TRBC sectors for different levels of market capitalization,
	we exclude portfolios that are formed only by five or less stocks.
	We perform the same analysis considering portfolios of stocks sorted for ten quantiles according to the ESG score.
	In Table \ref{table:constraint}, we report $E-\frac{A}{C} A_E$  and $G^*$ for the two different portfolio selection criteria (TRBC sectors and ESG score).
	In all the cases, $E-\frac{A}{C} A_E$ turns out to be negative, its absolute value increases as the capitalization of stocks increases.
	In the fourth and the fifth column of the table, we report the value of $G^*$ for the same portfolios. $G^*$ is consistently positive. These results show that the return-variance
	gain associated with an ESG mandate is borne out by the data.

	\begin{table}[]
		\centering
		\begin{tabular}{c|c c|c c}
			\toprule
			&\multicolumn{2}{c}{$E-\frac{A}{C} A_E$}&\multicolumn{2}{|c}{$G^*$}\\
			\% stocks	& TRBC sectors &ESG score & TRBC sectors &ESG score \\
			\midrule
			75\%& -0.27&-6.87&4.56\%&1.1\%\\
			70\%&-0.26 &-7.74&6.23\%&1.67\%\\
			60\% &-0.46&-6.98&3.77\%&2.25\%\\
			40\% &-1.19&-12.36&9.2\%&1.26\%\\
			20\% &-3.79&-15.47&4.11\%&0.05\%\\
			\bottomrule
		\end{tabular}
		\\[10pt]
		\caption{Value of $E-\frac{A}{C} A_E$ (second and third column) and of   $G^*$ (fourth and fifth column) for portfolios built through the TRBC sectors and considering 10 quantiles of the ESG scores. Portfolios include the top 75\%, 70\%, 60\%, 40\% and 20\% stocks sorted by market capitalization.}
		\label{table:constraint}
	\end{table}

	\section{ESG target over-performance}
	\label{EXT}
	We extend the model considering an ESG mandate with an over-performance target with respect to
	the benchmark as for the return.
	
	We derive the TEV ESG frontier when the ESG mandate also includes an ESG over-performance target $H\in \mathbb{R}$.
	As noticed in \cite{AVRAM}, norm-constrained institutions like pension funds are likely to have a positive $H$, other asset managers like hedge funds tend to have a negative $H$.
	A fund with a negative $H$ aims to outperform the benchmark and does not care of
	ESG considerations; a fund with a positive $H$ is a socially responsible player and targets a high ESG score.
	
	The optimization problem becomes:
	\begin{align}
		\min_{\mathbf{x}}&(\mathbf{x}-\mathbf{x_0})^{\top}\Omega (\mathbf{x}-\mathbf{x_0})\\
		&\text{subject to} \nonumber\\
		&\;\mathbf{x}^{\top}{\bf 1}=1 \\
		&(\mathbf{x}-\mathbf{x_0})^{\top}\boldsymbol{\xi} \ge H \label{CONST2*}
		\\
		&(\mathbf{x}-\mathbf{x_0})^{\top}\boldsymbol{\mu} =G.
	\end{align}

	The vector of portfolio weights of the TEV ESG frontier is
	\begin{equation}
		\mathbf{x}^*= \mathbf{x_0}-\frac{1}{2} \Omega^{-1}(\lambda_1 {\bf 1}+\lambda_2 \boldsymbol{\xi}+\lambda_3 \boldsymbol{\mu})
	\end{equation}
	
	where
	\begin{align}
		\label{LAGR*}
		\lambda_1&=   \frac{ 2 (E \, A_E - A \, B_E)G+(A\,E-A_EB)H }{ D_E} ,    \\
		\lambda_2&=  \frac{ 2 (A \, A_E - E \, C)G +DH}{ D_E} \le 0,  \label{eq:shad}  \\
		\lambda_3 &= \frac{ 2 (  B_E \, C-A_E^2)G +(A_E\,A-E\,C)H}{ D_E}
	\end{align}
	if the constraint (\ref{CONST2*}) is binding, and

	\[
	\mathbf{x}^*= \mathbf{x_0}-\frac{1}{2} \Omega^{-1}(\hat{\lambda}_1{\bf 1}+\hat{\lambda}_3  \boldsymbol{\mu})
	\]
	\begin{equation}
		\hat{\lambda}_1=   \frac{ 2 A \, G }{ D},     \
		\hat{\lambda}_3 = -\frac{ 2 C \, G }{ D},     \
		\label{eq:lagr_mult_roll*}
	\end{equation}
	if the constraint (\ref{CONST2*}) is not binding.
	
	The ESG constraint (\ref{CONST2*}) is binding in case
	\begin{equation}
		\label{POL}
		(E-\frac{A}{C}A_E)\frac{CG}{D}<H.
	\end{equation}
	
	We restrict our attention to the case $E-\frac{A}{C}A_E<0$, a condition that is borne out by the data.
	The ESG constraint (\ref{CONST2*}) is always binding if $G, \ H>0$, i.e., the asset manager has both a positive return and a positive ESG target. Instead, negative return-ESG targets ($G, \ H <0$) are never binding, and the mixed cases ($G>0, \ H<0$ and $G<0, \ H>0$) can lead to a binding or a non binding ESG constraint.

	The variance of the TEV ESG frontier is:
	\begin{align}
		&	Var_{TEV_{ESG}} (G) \\&= \begin{cases}
			\mathbf{x_0}^{\top}\Omega \mathbf{x_0}-   (\mathbf{x_0}^{\top} (\lambda_1 {\bf 1} + \lambda_2 \boldsymbol{\xi}   + \lambda_3  \boldsymbol{\mu})) + \frac{(A_E^2 - B_E C) G^2+(2EC-2AA_E)GH-DH^2}  {D_E}   &G\geq \frac{HD}{C(E-A/CA_E)}\\
			\mathbf{x_0}^{\top}\Omega \mathbf{x_{0}}-   ({\mathbf{x_{0}}}^{\top}( \hat{\lambda}_1 {\bf 1} +  \hat{\lambda}_3  \boldsymbol{\mu}))+\frac{CG^2}{D} &G<\frac{HD}{C(E-A/CA_E)}
		\end{cases}\;\;.
	\end{align}
	
	Similarly to the setting in Section \ref{FRO}, where there is no ESG over-performance target, the two frontiers intersect in $\frac{HD}{C(E-A/CA_E)}$ and
	$\hat{G}(H,\boldsymbol{\mu},\boldsymbol{\xi},\Omega,\mathbf{x_0})$ if
	$\hat{G}>\frac{HD}{C(E-A/CA_E)}$ or only in $\frac{HD}{C(E-A/CA_E)}$ if
	$\hat{G}\leq\frac{HD}{C(E-A/CA_E)}$. Only in the first case we observe a mean-variance improvement. Notice that, the case $H=0$ corresponds to what is discussed in  Section \ref{FRO} and, in that case, $\hat{G}=G^*$.
	In Figure \ref{figure:FrontierHss*}, we plot the three frontiers for different values of $H$. As $\hat{G}>\frac{HD}{C(E-\frac{A}{C}A_E)}$, in all the cases we observe a mean-variance improvement due to the ESG mandate.
	For all the examples we use the same assets as in Section \ref{PortAnalis}.

	\begin{figure}[!htb]
		\centering
		\begin{minipage}{.33\textwidth}
			\centering
			\includegraphics[width=1\linewidth, height=0.15\textheight]{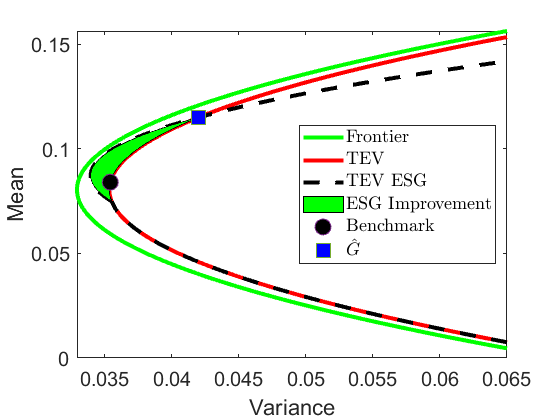}
		\end{minipage}%
		\begin{minipage}{0.33\textwidth}
			\centering
			\includegraphics[width=1\linewidth, height=0.15\textheight]{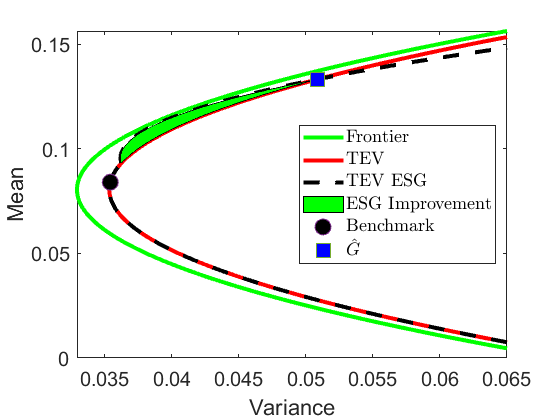}
		\end{minipage}
		\begin{minipage}{0.33\textwidth}
			\centering
			\includegraphics[width=1\linewidth, height=0.15\textheight]{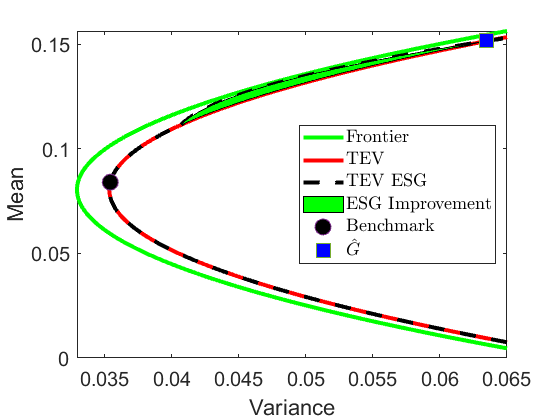}
		\end{minipage}
		\caption{The three frontiers when $E-A/CA_E<0$ and $\hat{G}>\frac{HD}{C(E-\frac{A}{C}A_E)}$, $H=0.1$ on the left, $H=-0.1$ in the middle and $H=-0.3$ on the right
			($\mathbf{x_0}^{\top}\boldsymbol{\mu}=8\%$, $\frac{A}{C}=8\%$).  \protect\label{figure:FrontierHss*}}
	\end{figure}
	
	As already discussed, condition $E-\frac{A}{C}A_E<0$ means that there is a negative ESG premium. In this setting, a return and an ESG over-performance
	target are in conflict with each other but there is space for a mean-variance improvement because a binding ESG constraint induces
	the asset manager to hold a portfolio with a higher M-SD ratio and therefore a lower variance.
	
	In Figure \ref{figure:FrontierHss*}, we observe that
	increasing the threshold of the ESG constraint (from the right to the left) there is more space for a mean-variance improvement.
	So a harsh ESG constraint is beneficial in a mean-variance perspective.  The result hinges on the mechanism described in Section \ref{PortAnalis}: the ESG constraint induces the asset manager to over-invest in assets with a high M-SD ratio. As $H$ is further increased ($\hat{G} \le \frac{HD}{C(E-A/CA_E)}$), the TEV ESG frontier is always dominated by the TEV frontier.
	In Figure \ref{figure:FrontierH*MV}, we report the M-SD ratio-ESG score combinations for portfolios belonging to the TEV frontier (red curve) and to the TEV ESG frontier (dashed black line) with $H=0.1$ on the left hand side, and $H=-0.3$ on the right side. We notice that in the first case the space for a M-SD improvement significantly increases.
		\begin{figure}[!htb]
		\centering
		\begin{minipage}{.5\textwidth}
			\centering
			\includegraphics[width=1\linewidth, height=0.25\textheight]{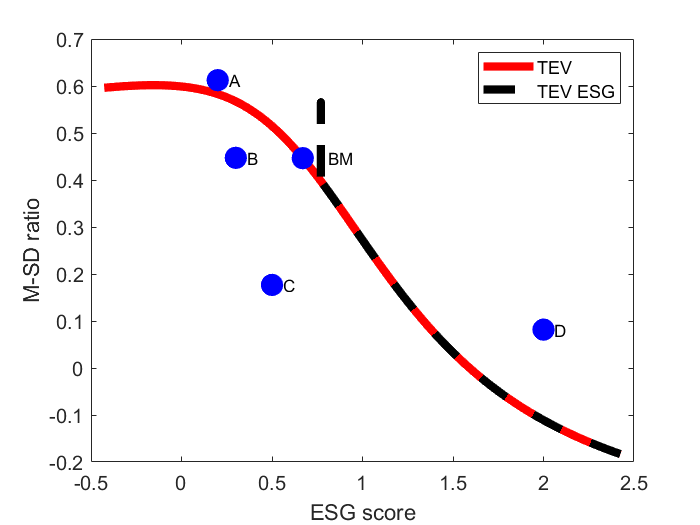}
		\end{minipage}%
		\begin{minipage}{0.5\textwidth}
			\centering
			\includegraphics[width=1\linewidth, height=0.25\textheight]{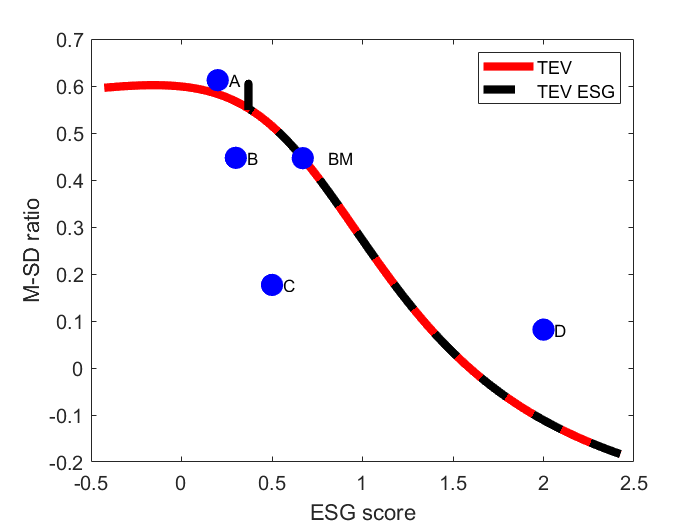}
		\end{minipage}
		\caption{	M-SD ratio-ESG score combinations for portfolios belonging to the TEV frontier (red curve) and to the TEV ESG frontier (dashed black line). $H=0.1$ on the left, and $H=-0.3$ on the right. BM is the benchmark.
		 \protect\label{figure:FrontierH*MV}}
	\end{figure}

	 A high ESG target leads to a risk reduction effect of an ESG mandate: an ESG responsible asset manager is able to reach a lower level of variance for a limited expected return over-performance target. Instead, a non-ESG responsible asset manager reaches a lower level of variance for a high expected return over-performance target.
When $E-\frac{A}{C}A_E<0$ and $\hat{G}>\frac{HD}{C(E-\frac{A}{C}A_E)}$,
	the mean-variance improvement allows to reduce the global minimum variance if $H>0$.

		We recall that $D>0$ and $D_E<0$, see Corollary \ref{cor:DE}. Then, if the ESG constraint is binding, its shadow price  $\lambda_2$ in \eqref{eq:shad} is linear and decreasing in both $H$ and $G$. In particular, the shadow price is zero if 	$(E-\frac{A}{C}A_E)\frac{CG}{D}=H$ and negative otherwise. Therefore, the cost of the ESG mandate increases both in $H$ and $G$.
	
	We can conclude that a socially responsible fund can achieve a significant ESG improvement for a small $G$, or even a negative $G$, while a hedge fund reaches a small improvement only for a large positive $G$.
	A positive ESG over-performance target acts as a risk reducer, instead a negative ESG over-performance target is a risk enhancer.
	
	\newpage
			
	\section{Conclusions}
	\label{CONC}
	This paper deals with the introduction of an ESG mandate for asset managers.
	We consider an asset manager who aims to minimize the tracking error variance and to maximize the expected
	over-performance with respect to the benchmark under the constraint that the ESG score of the portfolio is greater than that of the benchmark.
	
	We have explored the connection among frontiers derived by minimizing the portfolio variance, the TEV and adding an ESG constraint.
	Our findings show that ESG integration in asset management can mitigate inefficiencies of portfolios constructed minimizing TEV resulting in a smaller variance if there exists a negative relation between expected returns and ESG scores. For a moderate over-performance target, in our empirical analysis  between $0\%$ and $5\%$, an ESG mandate
	renders an improvement of the portfolio frontier in terms of mean-variance efficiency. Instead, for a high
	over-performance target, the ESG mandate leads to a
	worsening of the portfolio frontier. The result comes from the fact that
	a binding ESG mandate induces the asset manager
    to over-invest in assets with a high mean-standard deviation ratio, and therefore in assets with a low variance. 
	
	We have expanded the market equilibrium model with institutional investors to the case in which the fund manager
	is also subject to an ESG constraint.
	We prove that the market equilibrium, in the case of a binding  ESG mandate, implies a negative ESG premium.
	This negative ESG risk premium is consistently confirmed by the empirical evidence. In this framework, the negative ESG  premium arises from the sustainability constraint of institutional investors rather than as a remuneration of a risk factor.
    
	The two pieces of our analysis (mean-variance improvement and negative ESG premium in equilibrium) sustain each other: provided that a negative relation between expected returns and ESG scores holds true, then 	the ESG mandate is binding for institutional investors; in equilibrium with institutional investors
	dealing with a binding ESG mandate and retail investors, market returns are characterized by negative ESG premium confirming the hypothesis. The main insight is that an ESG mandate leads to a negative ESG premium but does not necessarily jeopardize fiduciary duties of asset managers.
	
	\section*{Acknowledgments}
	The authors are grateful to P. Angelini, E. Biffis, F. Cesarone, T. De Angelis, M. Fanari, M. Kacperczyk, A. Tarelli, and all participants to seminars at Università Cattolica del Sacro Cuore, Collegio Carlo Alberto, Università di Venezia, Bank of Italy and to the Quantitative Methods for Green Finance Workshop 2024, the GRASFI PhD Workshop, the XXV Quantitative Finance Workshop and the AMASES 2024 conference for their interesting comments and suggestions.
	
	This research has been developed within the MUSA—Multilayered Urban Sustainability Action—project, funded by the European Union—NextGenerationEU under the National Recovery and Resilience Plan (NRRP) Mission 4 Component 2 Investment Line 1.5: Strengthening of research structures and creation of R\&D “innovation ecosystems”, setup of “territorial leaders in R\&D”.
	Michele Azzone and Davide Stocco are members of the Gruppo Nazionale Calcolo
	Scientifico-Istituto Nazionale di Alta Matematica (GNCS-INdAM).

	\bibliography{sources}
	\bibliographystyle{tandfx}

	\appendix
\section{Technical Appendix}
\label{APP}
In what follows, we present the proofs of the main theoretical results.

	\begin{proposition}
	\label{prop:frontier}
	The vector $\mathbf{x}^*$ of portfolio weights of the TEV ESG frontier is
	\begin{equation}
		\mathbf{x}^*= \mathbf{x_0}-\frac{1}{2} \Omega^{-1}(\lambda_1 {\bf 1}+\lambda_2 \boldsymbol{\xi}+\lambda_3 \boldsymbol{\mu})
	\end{equation}
	where
	\begin{equation}
		\label{LAGR**}
		\lambda_1=   \frac{ 2 (E \, A_E - A \, B_E)G }{ D_E} ,    \
		\lambda_2=  \frac{ 2 (A \, A_E - E \, C)G }{ D_E} \le 0,    \
		\lambda_3 = \frac{ 2 (  B_E \, C-A_E^2)G }{ D_E}     \
	\end{equation}
	if the constraint (\ref{CONST2}) is binding.

	If the constraint (\ref{CONST2}) is not binding, then the vector of portfolio weights is

	\begin{equation}
		\mathbf{x}^*= \mathbf{x_0}-\frac{1}{2} \Omega^{-1}(\hat{\lambda}_1{\bf 1}+\hat{\lambda}_3  \boldsymbol{\mu}) 	\label{COND2}
	\end{equation}
	
	\begin{equation}
		\hat{\lambda}_1=   \frac{ 2 A \, G }{ D},     \
		\hat{\lambda}_3 = -\frac{ 2 C \, G }{ D},     \
		\label{eq:lagr_mult_roll**}
	\end{equation}
	where
	$$
	D_E = -2A \, E  \, A_E + A_E^2 \, B + A^2\,  B_E + E^2 \, C - B \,B_E  \, C, \ D = B\, C - A^2
	$$ and
	$$
	A={\bf 1}^{\top}\Omega^{-1}\boldsymbol{\mu}, \  \  B=\boldsymbol{\mu}^{\top}\Omega^{-1}\boldsymbol{\mu}, \  \   C={\bf 1}^{\top}\Omega^{-1}{\bf 1},
	$$
	$$
	A_E={\bf 1}^{\top}\Omega^{-1}\boldsymbol{\xi}, \  \  B_E=\boldsymbol{\xi}^{\top}\Omega^{-1}\boldsymbol{\xi}, \  \   E=\boldsymbol{\xi}^{\top}\Omega^{-1}\boldsymbol{\mu} \;\;.
	$$
\end{proposition}

	\begin{proof}
	The optimization problem is quadratic, a unique solution exists as the linear subspace satisfying the constraints is not empty and
	${\Omega}$ is positive definite.
	
	If $G=0$, then the solution of the optimization problem is $\mathbf{x}^*=\mathbf{x}_0$. In what follows, we focus on $G\neq 0$.
	
	The Lagrangian associated to the problem is
	\begin{equation*}
		\mathcal{L}(\mathbf{x},\lambda_1, \lambda_2, \lambda_3)=(\mathbf{x}-\mathbf{x_0})^{\top}\Omega (\mathbf{x}-\mathbf{x_0})+\lambda_1(\mathbf{x}^{\top}{\bf 1}-1)+\lambda_2(\mathbf{x}-\mathbf{x}_0)^{\top}\boldsymbol{\xi}+ \lambda_3((\mathbf{x}-\mathbf{x_0})^{\top}\boldsymbol{\mu}-G),
	\end{equation*}
	where $\lambda_1$, $\lambda_2$ and $\lambda_3$ are the Lagrange multipliers. The gradient of the Lagrangian is
	\begin{align*}
		\frac{\partial{\mathcal{L}}}{\partial \lambda_1} &= \mathbf{x}^{\top}{\bf 1}-1\\
		\frac{\partial{\mathcal{L}}}{\partial \lambda_2}& = (\mathbf{x}-\mathbf{x_0})^{\top}\boldsymbol{\xi}
		\\
		\frac{\partial{\mathcal{L}}}{\partial \lambda_3} &=(\mathbf{x}-\mathbf{x_0})^{\top}\boldsymbol{\mu}-G
		\\
		\frac{\partial{\mathcal{L}}}{\partial \mathbf{x}} &= 2\Omega (\mathbf{x}-\mathbf{x_0})+\lambda_1 {\bf 1}+\lambda_2 \boldsymbol{\xi}+\lambda_3 \boldsymbol{\mu}.
	\end{align*}
	
	The Kuhn-Tucker conditions, see \cite{ROCKA}, yield that
	$$
	\lambda_2 \le 0, \ \ \
	\lambda_2 (\mathbf{x}-\mathbf{x_0})^{\top}\boldsymbol{\xi}=0.
	$$
	
	Let us assume that the ESG constraint (\ref{CONST2}) is binding, then $\lambda_2 \le 0$. Setting $\frac{\partial{\mathcal{L}}}{\partial \mathbf{x}}=0$, we get the portfolio:
	\begin{equation*}
		\mathbf{x}^*= \mathbf{x_0}-\frac{1}{2} \Omega^{-1}(\lambda_1{\bf 1}+\lambda_2\boldsymbol{{\bf \xi}} +\lambda_3\boldsymbol{{\bf \mu}}).
	\end{equation*}
	Substituting $\mathbf{x}^*$ in (\ref{CONST2})-(\ref{CONST3}), we obtain the Lagrange multipliers in (\ref{LAGR**}).
	
	If the constraint (\ref{CONST2}) is not binding, then $\lambda_2=0$ and the minimization problem becomes:
	
	\begin{align}
		\min_{\mathbf{x}} &\,(\mathbf{x}-\mathbf{x_0})^{\top}\Omega \,(\mathbf{x}-\mathbf{x_0}) \label{eq:unconstr}\\
		&\text{subject to} \nonumber\\
		&\;\mathbf{x}^{\top}{\bf 1}=1 \\
		&(\mathbf{x}-\mathbf{x_0})^{\top}\boldsymbol{\mu} =G.
	\end{align}
	
	The Lagrangian of the problem is
	$$
	\mathcal{L}(\mathbf{x},\hat{\lambda}_1, \hat{\lambda}_3)=(\mathbf{x}-\mathbf{x_0})^{\top}\Omega (\mathbf{x}-\mathbf{x_0})+\hat{\lambda}_1(\mathbf{x}^{\top}{\bf 1}-1)+
	\hat{\lambda}_3((\mathbf{x}-\mathbf{x_0})^{\top}\boldsymbol{\mu}-G).
	$$
	Setting $\frac{\partial{\mathcal{L}}}{\partial \mathbf{x}}=0$,
	the Lagrange multipliers in (\ref{eq:lagr_mult_roll**}) are obtained and the optimal solution becomes:
	\begin{equation*}
		\mathbf{x}^*= \mathbf{x_0}-\frac{1}{2}\Omega^{-1}\left(\hat{\lambda}_1\mathbf{1}+\hat{\lambda}_3\boldsymbol{\mu}\right)
		\qedhere
	\end{equation*}
	
\end{proof}
\begin{corollary}\label{cor:DE}
	For $G=0$, the ESG constraint (\ref{CONST2}) is always binding. For $G\neq 0$, it is binding in case
	\begin{equation}
		\label{CONSTR2u}
		\begin{cases}
			E-\frac{A}{C}A_E<0,\quad &G>0\\
			E-\frac{A}{C}A_E>0, \quad &G<0\;\;.
		\end{cases}
	\end{equation}
	If the constraint is binding, then $ D_E<0$.
\end{corollary}

\begin{proof}
	
	If the ESG constraint (\ref{CONST2}) is binding, then the solution of the unconstrained optimization problem in (\ref{eq:unconstr}) does not satisfy the constraint.
	By substituting the unconstrained solution (\ref{COND}) in (\ref{CONST2}) and checking when the inequality  is not verified, we get the condition for a binding ESG constraint:
	\[
	( C \boldsymbol{\mu} -A\mathbf{1})^{\top}\Omega^{-1}\boldsymbol{\xi}\frac{G}{D} <0\;\;.
	\]
	Rearranging the terms and multiplying by $D$, we obtain
	\begin{equation}
		G(CE-AA_E)<0\;\;. \label{eq:bind}
	\end{equation}
	Dividing by $C >0$ and   $G\neq 0$, we get the condition in (\ref{CONSTR}).
	
	By the Kuhn-Tucker conditions, we have
	\begin{equation}
		\lambda_2=\frac{ 2 (A \, A_E - E \, C)G }{ D_E} \le 0\; \label{eq:KKT}
	\end{equation}
	if the constraint (\ref{CONST2}) is binding and therefore conditions (\ref{eq:bind}) and \eqref{eq:KKT}  imply $D_E<0$.
\end{proof}

}

\begin{proposition}\label{prop:eq_inv}
	The optimal portfolio for the institutional investor is
	\begin{equation}
		\mathbf{x_i}^*= \mathbf{x_{0i}}+\frac{1}{a_i} {\Omega}^{-1}( {\boldsymbol{\mu}}-\omega_1\mathbf{1}-\omega_2\boldsymbol{\xi}) \label{x_inst2}\;\;,
	\end{equation}
	where \begin{equation*}
		\begin{cases}
			\omega_1=\frac{A}{C}-\omega_2 Z,\quad \omega_2=\frac{\left(\boldsymbol{\mu}-\frac{A}{C}\mathbf{1}\right)^{\top}\Omega^{-1}\boldsymbol{\xi}}{\left(\boldsymbol{\xi}-Z\mathbf{1}\right)^{\top}\Omega^{-1}\boldsymbol{\xi}}\quad \quad &\text{if (\ref{CONSTR2}) is binding}\\
			\omega_1=\frac{A}{C}, \quad \omega_2=0 \quad \quad &\text{otherwise}
		\end{cases}
	\end{equation*}
	and $Z = \frac{\boldsymbol{\xi}^{\top}{\Omega}^{-1}\mathbf{1}}{\mathbf{1}^{\top}{\Omega}^{-1}\mathbf{1}}$.
	
	The ESG constraint  (\ref{CONSTR2}) is binding if and only if
	\begin{equation}
		\label{COND4**}
		E-\frac{A}{C}A_E<0\;\;.
	\end{equation}
\end{proposition}

\begin{proof}
	Let us assume that the constraint (\ref{CONSTR2}) is binding.
	
	In this case the Lagrangian is
	\[
	\mathcal{L}(\mathbf{x_i},\omega_1, \omega_2)=
	(\mathbf{x_i}-\mathbf{x_{0i}})^{\top}\boldsymbol{\mu}-\frac{a_i}{2} (\mathbf{x_i}-\mathbf{x_{0i}})^{\top}\Omega(\mathbf{x_i}-\mathbf{x_{0i}})+\omega_1(\mathbf{x_i}^{\top}\,\mathbf{1}-1)+\omega_2(\mathbf{x_i}^{\top}\boldsymbol{\xi}-\mathbf{x_{0i}}^{\top}\boldsymbol{\xi})\;\;,\]
	where $\omega_1$ and $\omega_2$ are the Lagrange multiplier. First order conditions render
	\begin{equation*}
		\mathbf{x_i}^*= \mathbf{x_{0i}}+\frac{1}{a_i} {\Omega}^{-1}(\boldsymbol {\mu}-\omega_1\mathbf{1}-\omega_2\boldsymbol{\xi}) \;\;.
	\end{equation*}
	Substituting $\mathbf{x_i}^*$ in the constraints, we get
	\begin{align*}
		\left( \mathbf{x_{0i}}+\frac{1}{a_i} {\Omega}^{-1}(\boldsymbol{\mu}-\omega_1\mathbf{1}-\omega_2\boldsymbol{\xi})\right)^{\top}\mathbf{1}&=1\\
		\left( \mathbf{x_{0i}}+\frac{1}{a_i} {\Omega}^{-1}(\boldsymbol{\mu}-\omega_1\mathbf{1}-\omega_2\boldsymbol{\xi})\right)^{\top}\boldsymbol{\xi}&= \mathbf{x_{0i}}^{\top}\boldsymbol {\xi}\;\;.
	\end{align*}
	Solving for $\omega_1$ and $\omega_2$, we obtain
	\begin{align*}
		\omega_1&= \frac{(\boldsymbol {\mu}-\omega_2\boldsymbol {\xi})^{\top}{\Omega}^{-1}\mathbf{1}}{\mathbf{1}^{\top}{\Omega}^{-1}\mathbf{1}}=\frac{A}{C}-\omega_2Z\\
		\omega_2&= \frac{\boldsymbol{\mu}^{\top}\Omega^{-1}\boldsymbol{\xi}-\frac{\boldsymbol{\mu}^{\top}\Omega^{-1}\mathbf{1}}{\mathbf{1}^{\top}\Omega^{-1}\mathbf{1}}\mathbf{1}^{\top}\Omega^{-1}\mathbf{\xi}}{\boldsymbol {\xi}^{\top}\Omega^{-1}\boldsymbol{\xi}-\mathbf{1}^{\top}\Omega^{-1}\boldsymbol{\xi}\frac{\boldsymbol{\xi}^{\top}\Omega^{-1}\mathbf{1}}{\boldsymbol{1}^{\top}\Omega^{-1}\mathbf{1}}}=\frac{\left(\boldsymbol{\mu}-\frac{A}{C}\mathbf{1}\right)^{\top}\Omega^{-1}\boldsymbol{\xi}}{\left(\boldsymbol{\xi}-Z\mathbf{1}\right)^{\top}\Omega^{-1}\boldsymbol{\xi}}\;\;.
	\end{align*}
	
	If the ESG constraint is not binding, then the Lagrangian becomes
	\[
	\mathcal{L}(\mathbf{x_i},\omega_1)=(\mathbf{x_i}-\mathbf{x_{0i}})^{\top}\boldsymbol{\mu}-\frac{a_i}{2} (\mathbf{x_i}-\mathbf{x_{0i}})^{\top}\Omega(\mathbf{x_i}-\mathbf{x_{0i}})+\omega_1(\mathbf{x_i}^{\top}\,\mathbf{1}-1).\]
	First order conditions render
	\begin{equation}
		\label{CANDID}
		\mathbf{x_i}^*= \mathbf{x_{0i}}+\frac{1}{a_i} {\Omega}^{-1}(\boldsymbol {\mu}-\omega_1\mathbf{1})\;\;.
	\end{equation}
	Substituting $	\mathbf{x_i}^*$ in (\ref{CONSTR2}) we get
	\begin{align*}
		\left( \mathbf{x_{0i}}+\frac{1}{a_i} {\Omega}^{-1}(\boldsymbol{\mu}-\omega_1\mathbf{1})\right)^{\top}\mathbf{1}&=1\;\;.
	\end{align*}
	yielding $\omega_1=\frac{A}{C}$.
	
	The ESG constraint is binding if
	the optimal portfolio in (\ref{CANDID}) does not satisfy the constraint (\ref{CONSTR2}),
	i.e., $\mathbf{x_i}^{*\top}\boldsymbol{\xi}<\mathbf{x_{0i}}^{\top}\boldsymbol{\xi}$:
	\begin{equation}
		\left( \mathbf{x_{0i}}+\frac{1}{a_i} {\Omega}^{-1}(\boldsymbol{\mu}-\frac{A}{C}\mathbf{1})\right)^{\top}\boldsymbol{\xi}< \mathbf{x_{0i}}^{\top}\boldsymbol {\xi}\;\;.
	\end{equation}
	Rearranging the terms, and multiplying by $a_i$, we get condition (\ref{COND4**}).
\end{proof}

\begin{proposition}\label{prop:equilibrium_tot}
	In equilibrium, if the ESG constraint is binding, then
	\begin{equation}
		\boldsymbol{\mu}={r_f^*}\mathbf{1}+\theta_1 \Omega \mathbf{x_m}-\theta_2\Omega \mathbf{x_0}-{\Gamma}\boldsymbol{\xi}\;\;, \label{eq:Eq2}
	\end{equation}
	where $r_f^*$, $\theta_1$, $\theta_2$ and $\Gamma$ are constants defined as follows:
	
	\begin{itemize}
		\item[i.] $r_f^*= \frac{\left(\omega_1\sum_{i=1}^I \frac{W_i}{a_i}+\sum_{l=1}^L \frac{W_l}{a_l}r_f\right)}{\delta}$
		\item[ii.] $\theta_1 = \frac{W_m}{\delta}$
		\item[iii.] $\theta_2 = \frac{ \sum_{i=1}^I W_i}{\delta}$
		\item[iv.] $\Gamma = -\frac{\sum_{i=1}^I\frac{W_i}{a_i}\omega_2}{\delta}$
	\end{itemize}
	
	and $\delta = \sum_{i=1}^I \frac{W_i}{a_i}+ \sum_{l=1}^L \frac{W_l}{a_l}$.
\end{proposition}

\begin{proof}
	By substituting (\ref{x_inst}) and (\ref{x_ind}) into (\ref{EQUI}), we obtain
	\begin{align}
		\boldsymbol{\mu}&=\frac{W_m\Omega \mathbf{x_m}+\left(\omega_1\sum_{i=1}^I \frac{W_i}{a_i}+\sum_{l=1}^L \frac{W_l}{a_l}r_f\right)\,\mathbf{1}-\sum_{i=1}^I W_i\Omega \mathbf{\,x_0}+\sum_{i=1}^I\frac{W_i}{a_i}\omega_2\boldsymbol{\xi}}{\sum_{i=1}^I \frac{W_i}{a_i}+ \sum_{l=1}^L \frac{W_l}{a_l}}\\
		&=:{r_f^*}\mathbf{1}+\theta_1 \Omega \mathbf{x_m}-\theta_2\Omega \mathbf{x_0}-{\Gamma}\boldsymbol{\xi},
	\end{align}
	where $\mathbf{x_0}=\sum_{i=1}^I {W_i}\mathbf{x_{0i}}$ is the aggregate benchmark portfolio of institutional investors.
\end{proof}


\end{document}